\documentclass[iop,twocolappendix]{emulateapj}
\usepackage{apjfonts}
\usepackage{amsmath,amssymb}
\renewcommand{\vec}[1]{\mathbf{#1}}
\newcommand{\diffp}[2]{\frac{\partial #1}{\partial #2}}

\shorttitle{Self-induced Strahl Scattering}
\shortauthors{Verscharen et al.}

\begin{document}

\title{Self-induced Scattering of Strahl Electrons in the Solar Wind}


\author{Daniel Verscharen$^{1,2}$, Benjamin D.~G.~Chandran$^{2,3}$, Seong-Yeop Jeong$^1$,\\ Chadi S.~Salem$^4$, Marc P.~Pulupa$^4$, and Stuart D.~Bale$^{4,5,6}$}
\affil{$^1$ Mullard Space Science Laboratory, University College London, Dorking, RH5~6NT, UK; d.verscharen@ucl.ac.uk, s.jeong.17@ucl.ac.uk\\
$^2$ Space Science Center, University of New Hampshire, Durham, NH 03824, USA; benjamin.chandran@unh.edu\\
$^3$ Department of Physics and Astronomy, University of New Hampshire, Durham, NH 03824, USA\\
$^4$ Space Sciences Laboratory, University of California, Berkeley, CA 94720, USA; salem@ssl.berkeley.edu, pulupa@ssl.berkeley.edu, bale@berkeley.edu\\
$^5$ Physics Department, University of California, Berkeley, CA 94720, USA\\
$^6$ The Blackett Laboratory, Imperial College London, SW7~2AZ, UK
}

\journalinfo{The Astrophysical Journal, 886:136 (11pp), 2019 December 1}
\submitted{Received 2019 June 6; revised 2019 October 5; accepted 2019 October 7; published 2019 November 29}

\begin{abstract}
We investigate the scattering of strahl electrons by microinstabilities as a mechanism for creating the electron halo in the solar wind. We develop a mathematical framework for the description of electron-driven microinstabilities and discuss the associated physical mechanisms. We find that an instability of the oblique fast-magnetosonic/whistler (FM/W) mode is the best candidate for a microinstability that scatters strahl electrons into the halo. We derive approximate analytic expressions for the FM/W instability threshold in two different $\beta_{\mathrm c}$ regimes,  where $\beta_{\mathrm c}$ is the ratio of the core electrons' thermal pressure to the magnetic pressure, and confirm the accuracy of these thresholds through comparison with numerical solutions to the hot-plasma dispersion relation. We find that the strahl-driven oblique FM/W instability creates copious FM/W waves under low-$\beta_{\mathrm c}$ conditions  when $U_{0\mathrm s}\gtrsim 3w_{\mathrm c}$, where $U_{0\mathrm s}$ is the strahl speed and $w_{\mathrm c}$ is the thermal speed of the core electrons. These waves have a frequency of about half the local electron gyrofrequency. We also derive an analytic expression for the oblique FM/W instability for $\beta_{\mathrm c}\sim 1$. The comparison of our theoretical results with data from the \emph{Wind} spacecraft confirms the relevance of the oblique FM/W instability for the  solar wind. The whistler heat-flux, ion-acoustic heat-flux, kinetic-Alfv\'en-wave heat-flux, and electrostatic electron-beam instabilities cannot fulfill the requirements for self-induced scattering of strahl electrons into the halo. We make predictions for the electron strahl close to the Sun, which will be tested by measurements from \emph{Parker Solar Probe} and \emph{Solar Orbiter}.
\end{abstract}

\keywords{instabilities -- plasmas -- solar wind -- Sun: corona -- turbulence -- waves}

\section{Introduction}

The solar wind is a plasma consisting of electrons, protons, and other ion species. Since the mass of an electron $m_{\mathrm e}$ is by about a factor of 1836 smaller than the mass of a proton $m_{\mathrm p}$, the electron contribution to the momentum flux in the solar wind is negligible. However, electrons are important to ensure quasi-neutrality, and their pressure gradient generates a substantial electrostatic field. Furthermore, the skewness provided by superthermal features in their distribution function supplies the solar wind with a significant heat flux \citep{gary99,scime99,pagel05,marsch06,verscharen19}.

Observations show that typical solar-wind electron distribution functions consist of three main populations: a thermal core, a superthermal halo, and a  field-aligned beam, which is usually called the \emph{strahl}  \citep[German for \emph{beam};][]{feldman75,rosenbauer77,pilipp87, pilipp87b,hammond96,fitzenreiter98,lin98,maksimovic00,gosling01,salem03b,wilson18}.  The thermal core typically exhibits  temperatures comparable to the proton temperature and contains about 95\% of the electrons. The halo is a tail in the distribution function extending to large velocities, which can be well modeled by a $\kappa$-distribution  \citep{maksimovic97,maksimovic05}. The halo is present in all directions with respect to the field; however, relative drifts and temperature anisotropies of both the halo and the core have been observed  \citep{stverak08,bale13}.  The electron strahl forms a ``shoulder'' in the distribution function. Its bulk velocity is shifted with respect to the electron core either parallel or antiparallel to the magnetic field, and its radial velocity component is almost always greater than the core's radial velocity component.  The nonthermal features of the distribution function are more distinctive in the fast solar wind. This observation is attributed to the typically weaker collisionality of the fast solar wind \citep{scudder79,scudder79b,phillips90,lie97,landi03,salem03,gurgiolo17}.

When binary collisions among electrons are sufficiently rare, instabilities can reduce the skewness of the distribution function and thereby limit the heat flux \citep{hollweg74,gary75,gary75b,feldman76,lakhina77,ramani78,lazar11}. Candidates for such instabilities include the electromagnetic whistler heat-flux instability \citep{gary77,gary94,gary00,lazar13}, fan instabilities of the lower-hybrid mode \citep{lakhina79,omelchenko94,krafft05,krafft06,shevchenko10}, the kinetic-Alfv\'en-wave (KAW) heat-flux instability \citep{gary75}, and a family of electrostatic instabilities \citep{gary78,gary07,pavan13} including the ion-acoustic heat-flux instability \citep{gary79,detering05}. 

In strahl-driven instabilities, the instability thresholds for bulk speed of the strahl component, $U_{0\mathrm s}$, in the reference frame in which the protons are at rest scale approximately as the electron Alfv\'en speed $v_{\mathrm{Ae}}\equiv B_0/\sqrt{4\pi n_{0\mathrm e}m_{\mathrm e}}$, where $n_{0\mathrm e}$ is the equilibrium electron number density and $B_0$ is the background magnetic field. Since $v_{\mathrm{Ae}}$ decreases with heliocentric distance in the inner heliosphere, the thresholds also decrease. This situation, therefore, leads to a quasi-continuous excitation of unstable wave modes during the solar wind's passage through the inner heliosphere beyond the point at which $U_{0\mathrm s}$ reaches the instability threshold for the first time. The exact location of this point depends on the properties of the innermost heliosphere and is thus still unknown. The unstable wave modes cause pitch-angle scattering of electrons, which reduces the strahl heat flux while transferring strahl electrons into the halo.

All electron-driven instabilities generate waves with wavelengths between the electron and the ion inertial length scales and inject these unstable waves into the fluctuation spectrum between these scales. Therefore, the understanding of electron microinstabilities is important for the understanding of the nature of fluctuations in the ion-to-electron spectral range, which is still under debate \citep{alexandrova09,sahraoui10,alexandrova12,chen12,he12,sahraoui12,salem12}.

In Section~\ref{sect:scenario}, we present the scenario we have in mind for how the electron distribution function evolves as the electrons stream away from the Sun. Section~\ref{sect:framework} introduces the mathematical framework of the quasilinear theory of wave--particle interactions and a conceptual discussion of why oblique fast-magnetosonic/whistler (FM/W) waves are a promising candidate for scattering strahl electrons into the halo. In Section~\ref{sect:fmw}, we present analytic expressions for the instability threshold of oblique FM/W waves in two different $\beta_{\mathrm c}$ regimes. Section~\ref{sect:observations} compares our theoretical results with in-situ electron observations in the solar wind. In Section~\ref{sect:others}, we discuss other electron-heat-flux instabilities and relate them to our predictions. Section~\ref{sec:conc} summarizes and concludes our treatment.

\section{Radial Evolution of the Strahl and Halo}\label{sect:scenario}


Measurements of the core, halo, and strahl densities ($n_{\mathrm c}$, $n_{\mathrm h}$, and $n_{\mathrm s}$, respectively) at different heliocentric distances show that  $n_{\mathrm s}/n_{\mathrm e}$ decreases with distance, while $n_{\mathrm h}/n_{\mathrm e}$ increases, where $n_{\mathrm e}=n_{\mathrm c}+n_{\mathrm h}+n_{\mathrm s}$  \citep{maksimovic05,graham17}. The quantity $\left(n_{\mathrm h}+n_{\mathrm s}\right)/n_{\mathrm e}$, however, remains almost constant between 0.3~au and 2~au \citep{stverak09}. This observation is striking evidence for the notion that the halo and strahl populations are closely related to each other, and that the electron halo is the result of a scattering of strahl electrons into the halo.
These observations have given rise to the following paradigm for the radial evolution of superthermal electrons in the solar wind:
\begin{enumerate}
\item electrons are heated to superthermal energies in (or very close to) the solar corona \citep{vocks03,owens08};
\item magnetic-moment conservation in the expanding magnetic field focuses the superthermal particles into the strahl;
\item microinstabilities, which constrain the strahl to stable regions in parameter space, regulate the strahl and the associated heat flux and generate waves that scatter the strahl into the halo during the passage of the plasma through the heliosphere, restricting the strahl to stable regions of parameter space.
\end{enumerate}
In this article, we address point 3 of this scenario.

\section{Resonant Wave--Particle Interactions}\label{sect:framework}

We consider a plasma consisting of protons (index p), core electrons (index c), and strahl electrons (index s) with a background magnetic field of the form $\vec B_0=(0,0,B_0)$. We neglect the halo for the sake of simplicity. We perform all calculations in the reference frame that moves with the proton bulk velocity. The plasma  fulfills the quasi-neutrality condition,
\begin{equation}
n_{0\mathrm p}=n_{0\mathrm c}+n_{0\mathrm s},
\end{equation}
and carries no field-parallel currents on average:
\begin{equation}\label{zerocurrents}
n_{0\mathrm c}U_{0\mathrm c}+n_{0\mathrm s}U_{0\mathrm s}=0,
\end{equation}
where $n_{0j}$ and $U_{0j}$ are the equilibrium number density and the equilibrium bulk velocity of species $j$, respectively.

\subsection{Quasilinear Theory of Resonant Wave--Particle Interactions}\label{sect:math}

Quasilinear theory describes the evolution of a plasma under the effects of resonant wave--particle interactions. Prerequisites for the application of this description are small amplitudes and small growth or damping rates (i.e., $|\gamma_k| \ll |\omega_{k\mathrm r}|$) of the resonant waves, where $\omega_{k\mathrm r}$ is the real part of the frequency $\omega_k$ at wavevector $\vec k$, and $\gamma_k$ is the imaginary part. These assumptions imply that the background distribution function and the wave amplitudes change on a timescale that is much longer than the wave periods. We use a cylindrical coordinate system for the velocity with the components $v_{\perp}$  and $v_{\parallel}$ perpendicular and parallel to $\vec B_0$. In the same coordinate system, the wavevector components are given by $k_{\perp}$ and $k_{\parallel}$. We denote the azimuthal angle of the wavevector as $\phi$. 

Resonant particles of species $j$ diffuse in velocity space according to the equation \citep{stix92}
\begin{multline}\label{qldiff}
\diffp{f_{j}}{t}=\lim _{V\to \infty}\sum \limits_{n=-\infty}^{+\infty}\frac{q_{j}^2}{8 \pi^2m_{j}^2}\int \frac{1}{Vv_{\perp}}\hat Gv_{\perp}\\
\times \delta(\omega_{k\mathrm r}-k_{\parallel}v_{\parallel}-n\Omega_{j}) \left|\psi^{j,n}_k \right|^2\hat Gf_{j}\mathrm d^3k,
\end{multline}
where $f_{j}$ is the distribution function of species $j$,  
\begin{equation}\label{goperator}
\hat G\equiv\left(1-\frac{k_{\parallel}v_{\parallel}}{\omega_{k\mathrm r}}\right)\diffp{}{v_{\perp}}+\frac{k_{\parallel}v_{\perp}}{\omega_{k\mathrm r}}\diffp{}{v_{\parallel}},
\end{equation}
and
\begin{multline}\label{eq:psink}
\psi_k^{j,n}\equiv \frac{1}{\sqrt{2}}\left[E_{k,\mathrm r}e^{i\phi}J_{n+1}(x_{j})+E_{k,\mathrm l}e^{-i\phi}J_{n-1}(x_{j})\right]\\
+\frac{v_{\parallel}}{v_{\perp}}E_{kz}J_n(x_{j}).
\end{multline}
The argument of the $\nu$th-order Bessel function $J_{\nu}(x_j)$ is given by $x_{j}\equiv k_{\perp}v_{\perp}/\Omega_{j}$, and $\Omega_{j}\equiv q_{j}B_0/m_{j}c$ is the (signed) cyclotron frequency of species $j$, where $q_{j}$ and $m_{j}$ are the particle charge and mass. The quantities 
\begin{equation}
\vec E_k(\vec k,t)\equiv \int\limits _V \vec E(\vec x,t)e^{-i\vec k\cdot \vec x}\mathrm d^3 x
\end{equation}
and
\begin{equation}
\vec B_k(\vec k,t)\equiv \int\limits_V \vec B(\vec x,t)e^{-i\vec k\cdot \vec x}\mathrm d^3 x
\end{equation}
are the Fourier transforms of the electric and magnetic fields $\vec E(\vec x,t)$ and $\vec B(\vec x,t)$, after these fields have been multiplied by a window function of volume $V$.\footnote{This Fourier transform convention is described in greater detail by \citet{stix92}, although his definitions of $\vec E_k$ and $\vec B_k$ differ from ours by a factor of $\left(2\pi\right)^{-3/2}$.} The left and right circularly polarized components of the electric field are given by $E_{k,\mathrm l}\equiv (E_{kx}+iE_{ky})/\sqrt{2}$ and $E_{k,\mathrm r}\equiv(E_{kx}-iE_{ky})/\sqrt{2}$

When quasilinear diffusion occurs according to Equation~(\ref{qldiff}), particles diffuse in velocity space along curves of constant energy in the wave frame (i.e., the reference frame that moves with the speed $\omega_{k\mathrm r}/k_{\parallel}$ along $\vec B_0$). This diffusive flux of particles is locally tangent to semicircles centered on the parallel phase velocity $v_{\mathrm{ph}}\equiv \omega_{k\mathrm r}/k_{\parallel}$ (see also Figure~\ref{fig_diff_paths_cases}), which satisfy the equation
\begin{equation}\label{qlcircle}
\left(v_{\parallel}-v_{\mathrm{ph}}\right)^2+v_{\perp}^2=\mathrm{constant}.
\end{equation} 
At the same time, Equation~(\ref{qldiff}) allows for diffusion only from higher phase-space densities to lower phase-space densities. Only waves and particles fulfilling the resonance condition
\begin{equation}\label{rescond}
\omega_{k\mathrm r}=k_{\parallel}v_{\parallel}+n\Omega_{j}
\end{equation}
participate in the resonant wave--particle interaction due to the $\delta$-function in Equation~(\ref{qldiff}).

Complementary to Equation~(\ref{qldiff}), \citet{kennel67} calculated the growth/damping rate of waves with $|\gamma_k|\ll |\omega_{k\mathrm r}|$ in quasilinear theory and found that the contribution of  species $j$ to $\gamma_k$ is given by
 \begin{equation}\label{total_growthrate}
\gamma_k^{j}=\sum\limits_{n=-\infty}^{+\infty} \gamma_k^{j,n},
 \end{equation}
where
\begin{multline}\label{growthrate}
\frac{\gamma_k^{j,n}}{|\omega_{k\mathrm r}|}=\frac{\pi}{8n_{0j}} \left|\frac{\omega_{k\mathrm r}}{k_{\parallel}} \right|\left(\frac{\omega_{\mathrm pj}}{\omega_{k\mathrm r}}\right)^2 \int \limits_0^{\infty}\mathrm dv_{\perp} v_{\perp}^2\\ 
\times \int \limits_{-\infty}^{+\infty}\mathrm dv_{\parallel}\,\delta\left(v_{\parallel}-\frac{\omega_{k\mathrm r}-n\Omega_{j}}{k_{\parallel}}\right)\frac{\left|\psi_k^{j,n}\right|^2 \hat Gf_{0j}}{W_k},
\end{multline}
\begin{equation}\label{waveenerg}
W_k\equiv \frac{1}{16\pi}\left.\left[\vec B^{\ast}_{k}\cdot\vec B_{k}+\vec E^{\ast}_{k}\cdot \diffp{}{\omega}(\omega \varepsilon_{\mathrm h})\vec E_{k}\right]\right|_{\omega=\omega_{k\mathrm r}},
\end{equation}
$\varepsilon_{\mathrm h}$ denotes the Hermitian part of the dielectric tensor, and $f_{0j}$ is the background distribution function of species $j$.\footnote{For the following discussion, we focus on waves with $W_k>0$. The arguments are inverted for negative-energy waves. This effect is, however, not relevant for the parameter range explored in this study \citep[cf.][]{verscharen13}.} The plasma frequency of species $j$ is defined by $\omega_{\mathrm pj}\equiv \sqrt{4\pi n_{0j}q_{j}^2/m_{j}}$. We assume that the background distribution functions $f_{0j}$ are represented by drifting Maxwellians,
\begin{equation}\label{Maxwell}
f_{0j}=\frac{n_{0j}}{\pi^{3/2}w_{j}^3} \exp\left(-\frac{v_{\perp}^2+\left(v_{\parallel}-U_{0j}\right)^2}{w_{j }^2}\right),
\end{equation}
where $w_{j }\equiv \sqrt{2k_{\mathrm B}T_{0j}/m_{j}}$ is the thermal speed of species $j$, $k_{\mathrm B}$ is the Boltzmann constant, and $T_{0j}$ is the equilibrium temperature of species $j$. 
The set of Equations~(\ref{total_growthrate}) and (\ref{qldiff}) couple the evolution of the waves and particles in the presence of resonant wave--particle interactions and fulfill energy conservation \citep{kennel67,chandran10b}. 

\subsection{Conceptual Predictions of Quasilinear Diffusion}

\begin{figure}
\epsscale{0.8}
\plotone{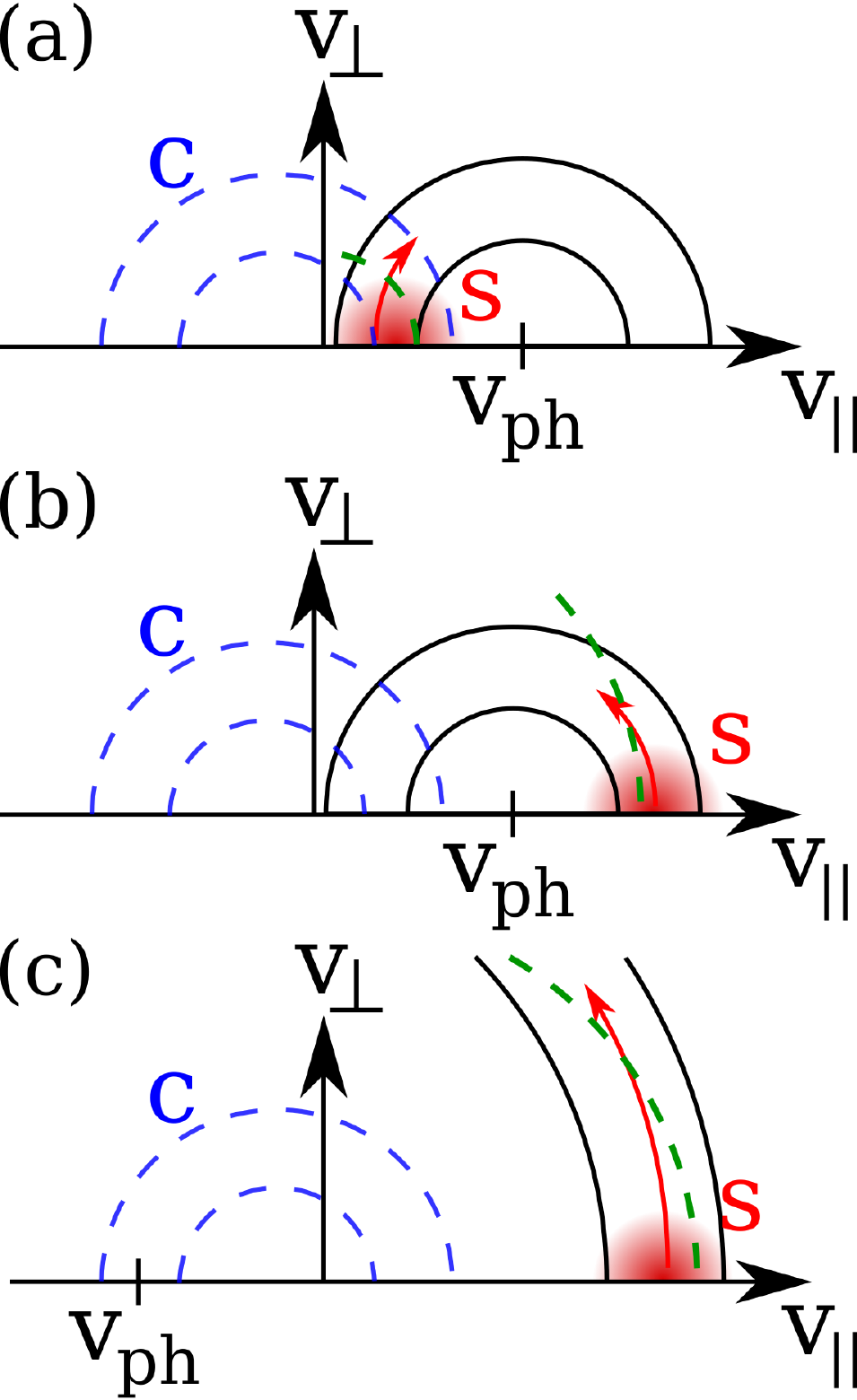}
\caption{Quasilinear diffusion paths in a core-strahl electron distribution function. The core distribution is shown as blue dashed semicircles and the strahl distribution as a red shaded semicircle. The particle diffusion path (red arrow) is locally tangent to semicircles (black) about the parallel phase speed $v_{\mathrm{ph}}\equiv \omega_{k\mathrm r}/k_{\parallel}$ and directed toward lower strahl phase-space density. The green circle shows a contour of constant kinetic energy (i.e., $v_{\perp}^2+v_{\parallel} ^2=\text{constant}$). Under typical solar-wind conditions, $w_{\mathrm c}\sim v_{\mathrm {Ae}}$, and core and strahl overlap in velocity space. (a) $0<U_{0\mathrm s}<v_{\mathrm{ph}}$; (b) $0<v_{\mathrm{ph}}<U_{0\mathrm s}$; (c) $v_{\mathrm{ph}}<0$ and $U_{0\mathrm s}>0$. }
\label{fig_diff_paths_cases}
\end{figure}

Figure~\ref{fig_diff_paths_cases} shows quasilinear diffusion paths for resonant strahl particles in velocity space. 
The relative alignment of the gradients of $f_{0j}$ in velocity space and the semicircles given by Equation~(\ref{qlcircle}) determines if the resonant particles lose or gain kinetic energy during the quasilinear diffusion process. 
In case (a), $0<U_{0\mathrm{s}}<v_{\mathrm{ph}}$, the resonant particles gain kinetic energy as they diffuse (i.e., their distance, $v_{\perp}^2+v_{\parallel}^2$, from the origin increases). They remove this energy from the resonant wave, which consequently decreases in amplitude. Therefore, case (a) represents a configuration that leads to wave damping.
In case (b), $0<v_{\mathrm{ph}}<U_{0\mathrm s}$, the resonant particles lose kinetic energy as they diffuse and transfer this energy to the resonant wave, which consequently grows in amplitude. Therefore, case (b) represents a configuration that leads to wave instability.
Case (c), in which $U_{0\mathrm s}>0$ and $v_{\mathrm{ph}}<0$, represents an additional configuration that leads to wave damping since the resonant particles gain kinetic energy during  quasilinear diffusion.
We, therefore, conclude that 
\begin{equation}\label{gencond}
0<v_{\mathrm{ph}}<U_{0\mathrm s}
\end{equation}
is a necessary condition in order for strahl electrons to lose kinetic energy through a wave--particle resonance.  At this point, we limit ourselves to $\omega_{k\mathrm r}>0$ and $k_{\parallel}>0$ without loss of generality.\footnote{Our arguments for instability also apply to a configuration in which $U_{0\mathrm s}<v_{\mathrm{ph}}<0$ \citep[see][]{verscharen13,verscharen13b}. If the strahl streams away from the Sun, the configuration $U_{0\mathrm s}<0$ corresponds to the case in which $\vec B_0\cdot \hat{\vec e}_r<0$, where $\hat{\vec e}_r$ is the radial unit vector in the solar rest frame. }

\begin{figure}
\epsscale{1.2}
\plotone{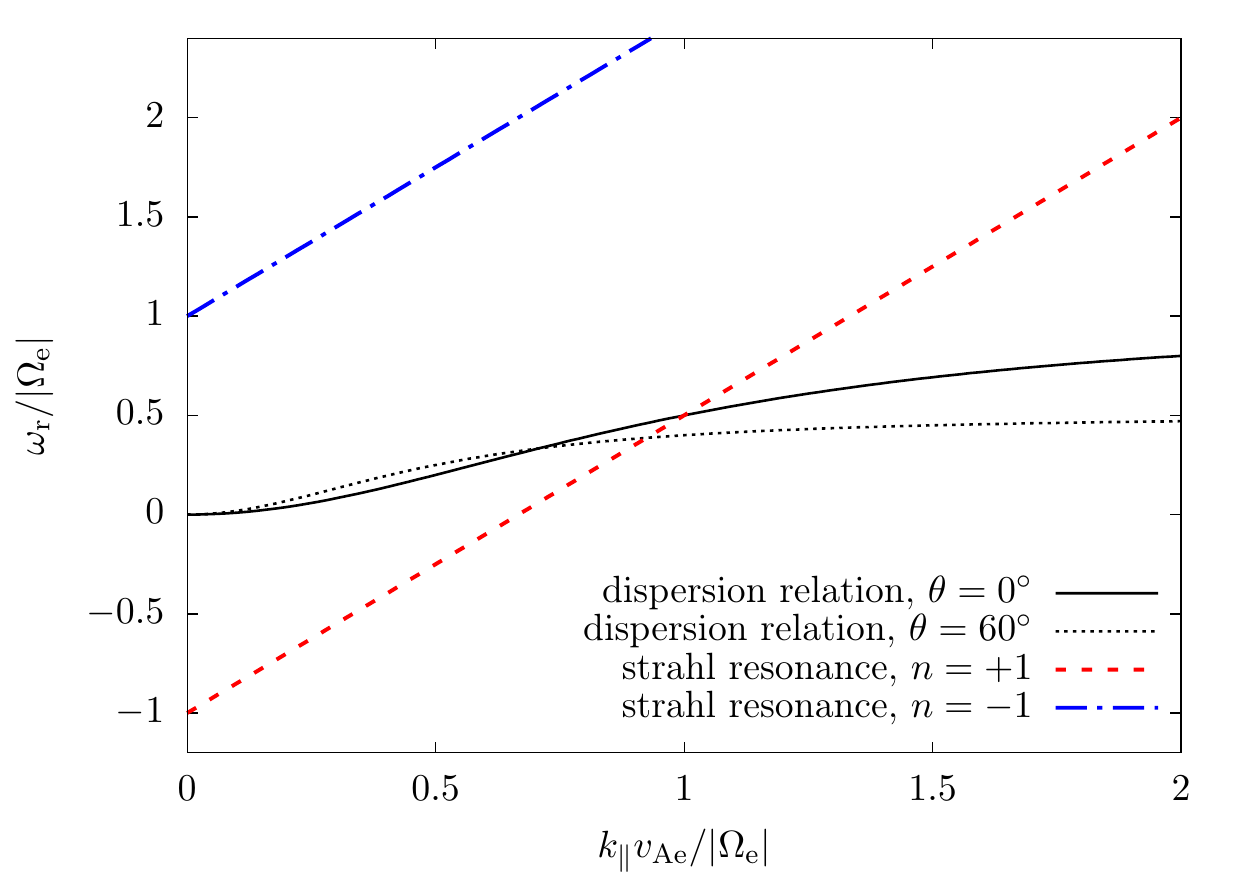}
\caption{Dispersion relation of the FM/W wave for $\theta=0^{\circ}$ (solid black) and $\theta=60^{\circ}$ (dashed black) from Equation~(\ref{approxDR}), as well as the strahl resonance conditions according to Equation~(\ref{rescond}) with $v_{\parallel}=1.5v_{\mathrm {Ae}}$ for $n=+1$ (red) and $n=-1$ (blue). Strahl electrons with $v_{\parallel}>0$ can only resonate with waves with $v_{\mathrm{ph}}>0$ through resonances with $n\ge 0$.}
\label{resonance_schematic}
\end{figure}

The frequency $\omega_{k\mathrm r}$ is associated with the wavevector through the linear dispersion relation.
Figure~\ref{resonance_schematic} shows two representative plots for the FM/W-wave dispersion relation (for details, see Section~\ref{sect:fmw}) with $\omega_{k\mathrm r}>0$ and $k_{\parallel}>0$. In addition, we plot the resonance condition in Equation~(\ref{rescond}) for $n=-1$ and $n=+1$. Resonant interactions only occur for waves and particles for which the line corresponding to the resonance condition intersects the plot of the dispersion relation.  Figure~\ref{resonance_schematic} illustrates that strahl electrons with $v_{\parallel}>0$ only fulfill the resonance condition with FM/W waves with $\omega_{k\mathrm r}>0$ and $k_{\parallel}>0$ through resonance with $n\ge 0$. They cannot fulfill the resonance condition with these waves through resonances with $n\le -1$ since $\omega_{k\mathrm r}<|\Omega_{\mathrm e}|$ for all $\vec k$ and $\Omega_{\mathrm e}<0$. We do not consider the $n=0$ strahl resonance since this resonance only excites instabilities when there is a \emph{bump-on-tail} distribution (i.e., $\partial f_{j}/\partial v_{\parallel}>0$ at the resonance speed), which is not observed (see also Section~\ref{elstat}). Moreover, instabilities driven via the $n=0$ strahl resonance are unable to account for the halo formation since they cause particles to diffuse only in $v_{\parallel}$ and not in $v_{\perp}$.

Although we will allow for oblique modes, we still limit ourselves to $k_{\perp}\rho_{\mathrm e}\ll 1$ to avoid cyclotron damping by the core, where $\rho_{\mathrm e}\equiv w_{\mathrm e}/|\Omega_{\mathrm e}|$, a point that we discuss further in Section~\ref{sect:criterion} below.

When $k_{\perp}\rho_{\mathrm e}\ll1$ and $n>0$, the only non-negligible term in Equation~(\ref{eq:psink}) is  $E_{k,\mathrm l}e^{-i\phi}J_{n-1}(x_j)$  which, moreover, is nonzero only when $n=+1$, because $J_{\nu}(x_j)\rightarrow 0$ for $x_j\rightarrow 0$ for all $\nu\neq 0$. Therefore, in order for electrons to undergo an $n>0$ resonance with FM/W waves with $k_{\perp}\rho_{\mathrm e}\ll 1$, the FM/W waves must have a left-circularly polarized component (i.e., $E_{k,\mathrm l}\neq 0$). This requirement rules out the parallel-propagating FM/W wave, which is purely right-circularly polarized (i.e., $E_{k,\mathrm l}=0$), forcing us to consider obliquely propagating FM/W waves.\footnote{The only left-circularly polarized parallel-propagating normal mode is the Alfv\'en/ion-cyclotron mode. This mode, however, has low frequencies ($\lesssim \Omega_{\mathrm p}$) compared to $|\Omega_{\mathrm e}|$. At these frequencies and under typical solar-wind conditions, thermal protons fulfill the resonance condition in Equation~(\ref{rescond}), so that the mode is prone to strong proton-cyclotron damping.}

Figure~\ref{fig_diff_paths_cases} (b) shows a case that satisfies all of these requirements for the instability of the oblique FM/W wave. The diffusing particles  increase in $v_{\perp}$ and (slightly) decrease in $v_{\parallel}$. These particles are the seed for the halo population. However, scattering by the initially excited FM/W waves does not fully describe the formation of the observed halo since it is restricted to particles in a certain range in $v_{\parallel}$ that fulfill the resonance condition. The scattered seed population, however, represents a strong deformation of the electron distribution function, which may eventually relax through secondary instabilities into a more symmetric halo about the electron core. A detailed study of the nonlinear evolution of the electron system is, however, beyond the scope of this work.


\section{Instability of the Oblique Fast-magnetosonic/Whistler Mode}\label{sect:fmw}

In this section, we derive approximate analytical expressions for the instability thresholds of the oblique FM/W mode in a plasma containing an electron strahl.  \citet{gary75} and \citet{gary75b} refer to this instability as the ``magnetosonic instability.'' An instability of the oblique FM/W mode has recently been discussed in the context of the solar wind \citep{horaites18,vasko19}. This instability is also a candidate to explain heat-flux regulation in other astrophysical plasmas such as the intracluster medium \citep{roberg16,roberg18} and in solar flares \citep{roberg19}.

\subsection{Instability Mechanism and Dispersion Relations}\label{sect:criterion}

We assume that the instability drive by strahl electrons is most efficient at the center of the strahl distribution function, i.e., by particles with $v_{\parallel}=U_{0\mathrm s}$ and $v_{\perp}=0$.  Through the $n=+1$ resonance, the oblique FM/W instability only occurs at frequencies of about
\begin{equation}\label{strahlcond}
\omega_{\mathrm r}=k_{\parallel}U_{0\mathrm s}-|\Omega_{\mathrm e}|
\end{equation}
according to Equation~(\ref{rescond}).

\begin{figure}
\epsscale{1.2}
\plotone{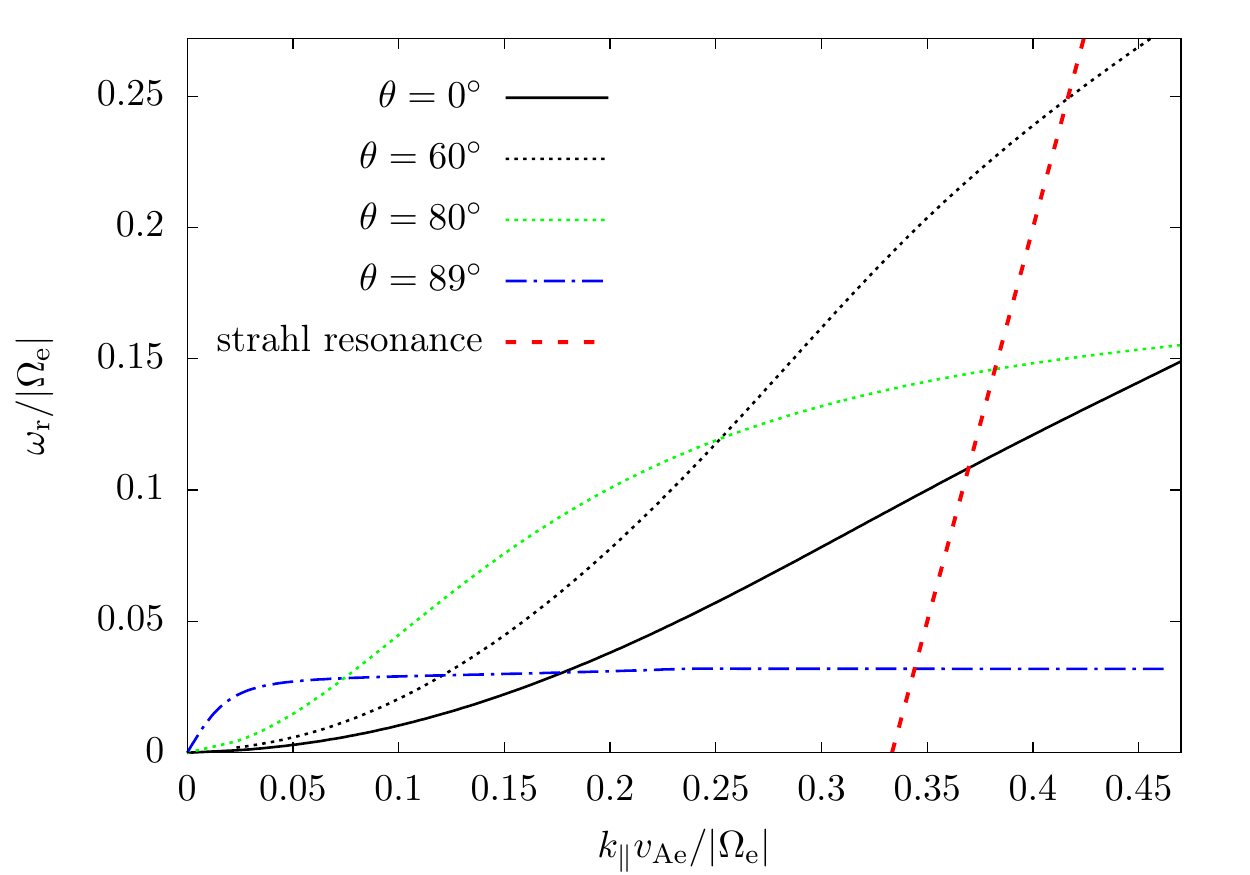}
\caption{Hot-plasma dispersion relations for the FM/W mode at different angles of propagation in an electron--proton plasma with $T_{0\mathrm p}=T_{0\mathrm e}$. We neglect the influence of the strahl on the dispersion relation. For the parallel-propagating mode and for the mode with $\theta=60^{\circ}$, we use $\beta_{\mathrm p}=1$. For the other two modes, $\beta_{\mathrm{p}}=0.001$ in order to avoid strong linear damping. The strahl-resonance line shows Equation~(\ref{strahlcond}) for $U_{0\mathrm s}=3v_{\mathrm{Ae}}$. }
\label{fig_dispersions}
\end{figure}

We discuss the properties of the FM/W mode at different angles of propagation in the $\omega_{\mathrm r}$-$k_{\parallel}$ plane in Figure~\ref{fig_dispersions}, where we show four solutions from the full dispersion relation of a hot electron--proton plasma. We use the linear Vlasov--Maxwell solver NHDS  \citep[for details on the numerics, see][]{verscharen13a,verscharen18b}.
In addition, we show Equation~(\ref{strahlcond}). The intersection between this line and a plot of the dispersion relation indicates a wavenumber and frequency at which the resonance condition between the wave and an electron with $v_{\parallel}=U_{0\mathrm s}$  is fulfilled.
Since highly oblique modes cease to exist in plasmas with Maxwellian background distributions at large $\beta_{\mathrm p}\equiv 8\pi n_{0\mathrm p}T_{0\mathrm p}/B_0^2$ due to increasing Landau damping, we apply lower $\beta_{\mathrm p}$ for the highly oblique cases in Figure~\ref{fig_dispersions}.\footnote{We note that the lower $\beta_{\mathrm p}$-value used for the highly oblique solutions in Figure~\ref{fig_dispersions} is not representative for the solar wind at 1\,au, but represents coronal conditions instead.} We note that $k_{\perp}$ in the case with $\theta=89^{\circ}$ is by a factor $\tan \theta\sim 57$ greater than $k_{\parallel}$. 
The FM/W mode exists in two regimes in the wavenumber range in which Equation~(\ref{strahlcond}) can be fulfilled---i.e., where the line of the strahl resonance intersects with the corresponding plot of the dispersion relation. Regime 1 is the \emph{whistler regime} in which the angle of propagation fulfills $0^{\circ}\le\theta \lesssim 70^{\circ}$, and
\begin{equation}\label{approxDR}
\omega_{k\mathrm r}\approx \frac{kk_{\parallel}v_{\mathrm{Ae}}^2}{|\Omega_{\mathrm e}|\left(1+k^2d_{\mathrm e}^2\right)},
\end{equation}
where $d_{\mathrm e}\equiv v_{\mathrm{Ae}}/|\Omega_{\mathrm e}|$ is the electron inertial length. Equation~(\ref{approxDR}) follows from the cold-plasma dispersion relation in a plasma with a single electron species under the assumption that $kd_{\mathrm p}\gg 1$, where $d_{\mathrm p}$ is the proton inertial length, and is approximately valid for small $\beta_{\mathrm c}$ and $\beta_{\mathrm p}$.
The dispersion relation asymptotes toward $\sim |\Omega_{\mathrm e}|\cos\theta$ for large $k_{\parallel}$ provided that $\cos^2\theta \gtrsim m_{\mathrm e}/m_{\mathrm p}$.  
In the highly oblique limit (regime 2; i.e.,  $\cos^2\theta \lesssim m_{\mathrm e}/m_{\mathrm p}$), the wave propagates in the \emph{lower-hybrid regime}. Its frequency asymptotes toward a frequency of order the lower-hybrid frequency,
\begin{equation}
\omega_{\mathrm{LH}}\equiv \frac{\omega_{\mathrm{pp}}}{\sqrt{1+\displaystyle \frac{\omega_{\mathrm{pe}}^2}{\Omega_{\mathrm e}^2}}},
\end{equation}
as long as thermal corrections are small \citep{verdon09}.

\subsection{Analytical Instability Thresholds}

To determine whether the oblique FM/W wave is unstable, we must consider not only the instability drive provided by the strahl, but also the possibility of damping by the core electrons. According to Equations~(\ref{rescond}) and (\ref{Maxwell}), cyclotron damping by the core with $n=-1$ occurs at wavenumbers and frequencies that fulfill
\begin{equation}\label{corecond}
-k_{\parallel}w_{\mathrm c}+|\Omega_{\mathrm e}| \lesssim \omega_{\mathrm r}\lesssim k_{\parallel} w_{\mathrm c}+|\Omega_{\mathrm e}|.
\end{equation}
Landau damping by the core with $n=0$ occurs at wavenumbers and frequencies that fulfill
\begin{equation}\label{coreLcond}
-k_{\parallel}w_{\mathrm c} \lesssim \omega_{\mathrm r}\lesssim k_{\parallel} w_{\mathrm c}.
\end{equation}
Strahl driving with $n=+1$ occurs at wavenumbers and frequencies that fulfill Equation~(\ref{strahlcond}).

\begin{figure}
\epsscale{1.2}
\plotone{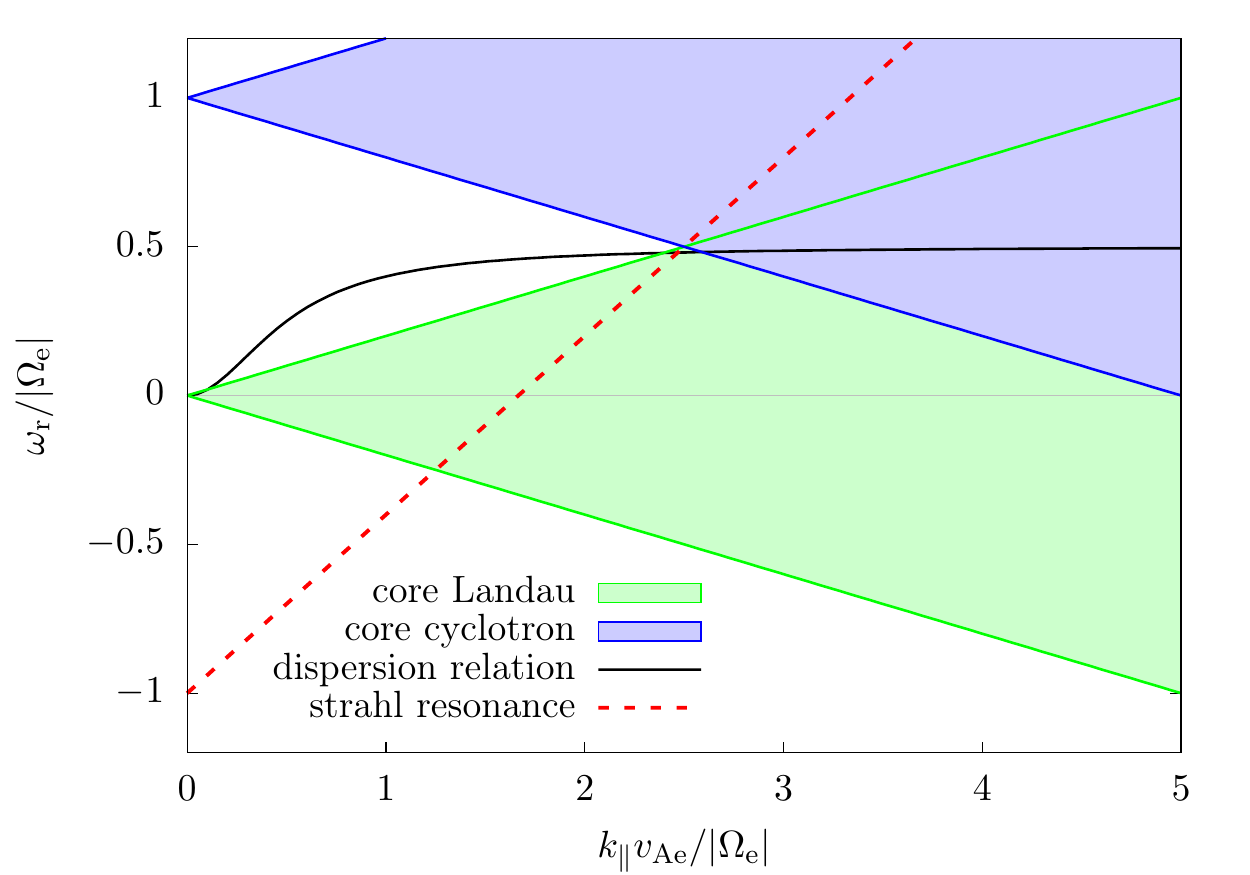}
\caption{Dispersion relation and resonance conditions for the FM/W mode with $\theta=60^{\circ}$ in the low-$\beta_{\mathrm c}$ case. The black line shows Equation~(\ref{approxDR}). The blue and green areas show Equations~(\ref{corecond}) and (\ref{coreLcond}), respectively, and the red line shows Equation~(\ref{strahlcond}) with $U_{0\mathrm s}=3w_{\mathrm c}$. We use $w_{\mathrm c}=0.2v_{\mathrm{Ae}}$. This situation represents a marginally stable state for the oblique FM/W instability.}
\label{resonances1}
\end{figure}
Figure~\ref{resonances1} shows the plot of the dispersion relation from Equation~(\ref{approxDR}), the strahl-resonance line from Equation~(\ref{strahlcond}), and the parameter space in which core Landau damping and core cyclotron damping act from Equations~(\ref{corecond}) and (\ref{coreLcond}). For this plot, we have assumed that $w_{\mathrm c}\ll v_{\mathrm{Ae}}$. To a first approximation, the FM/W wave is unstable if there is a wavenumber range in which Equation~(\ref{strahlcond}) is fulfilled and Equations~(\ref{corecond}) and (\ref{coreLcond}) are not fulfilled.  The resonance line in Figure~\ref{resonances1} represents the minimum $U_{0\mathrm s}$ for which strahl driving can occur in a wavenumber range in which neither core cyclotron damping nor core Landau damping acts.  In order for the dispersion relation to intersect this resonance line within the white triangle, two conditions must be met: (1) $\beta_{\mathrm e}$ must be small (otherwise the dispersion relation will lie within the Landau-damped region -- see Figure~\ref{resonances1}), and (2) $\theta$ must be $\simeq 60^{\circ}$. When these conditions are satisfied, the frequency of the resonant waves is given by 
\begin{equation}\label{reg2freq}
\omega_{k\mathrm r}\approx \frac{1}{2}|\Omega_{\mathrm e}|.
\end{equation}
Furthermore, it follows from Equations~(\ref{corecond}) and (\ref{coreLcond}) that the wavenumber of the resonant waves satisfies
\begin{equation}\label{kcond}
k_{\parallel}=k_{\mathrm{crit}}\equiv \frac{1}{2}\frac{\left|\Omega_{\mathrm e}\right|}{w_{\mathrm c}},
\end{equation}
where $k_{\mathrm{crit}}$ is the minimum wavenumber at which the green and blue regions in Figure~\ref{resonances1} overlap; i.e., the minimum wavenumber at which both core Landau damping and core cyclotron damping can occur.
Combining Equations~(\ref{strahlcond}), (\ref{reg2freq}), and (\ref{kcond}) leads to the instability criterion
\begin{equation}\label{cond2}
U_{0\mathrm s}\gtrsim 3 w_{\mathrm c}.
\end{equation}

\begin{figure}
\epsscale{1.2}
\plotone{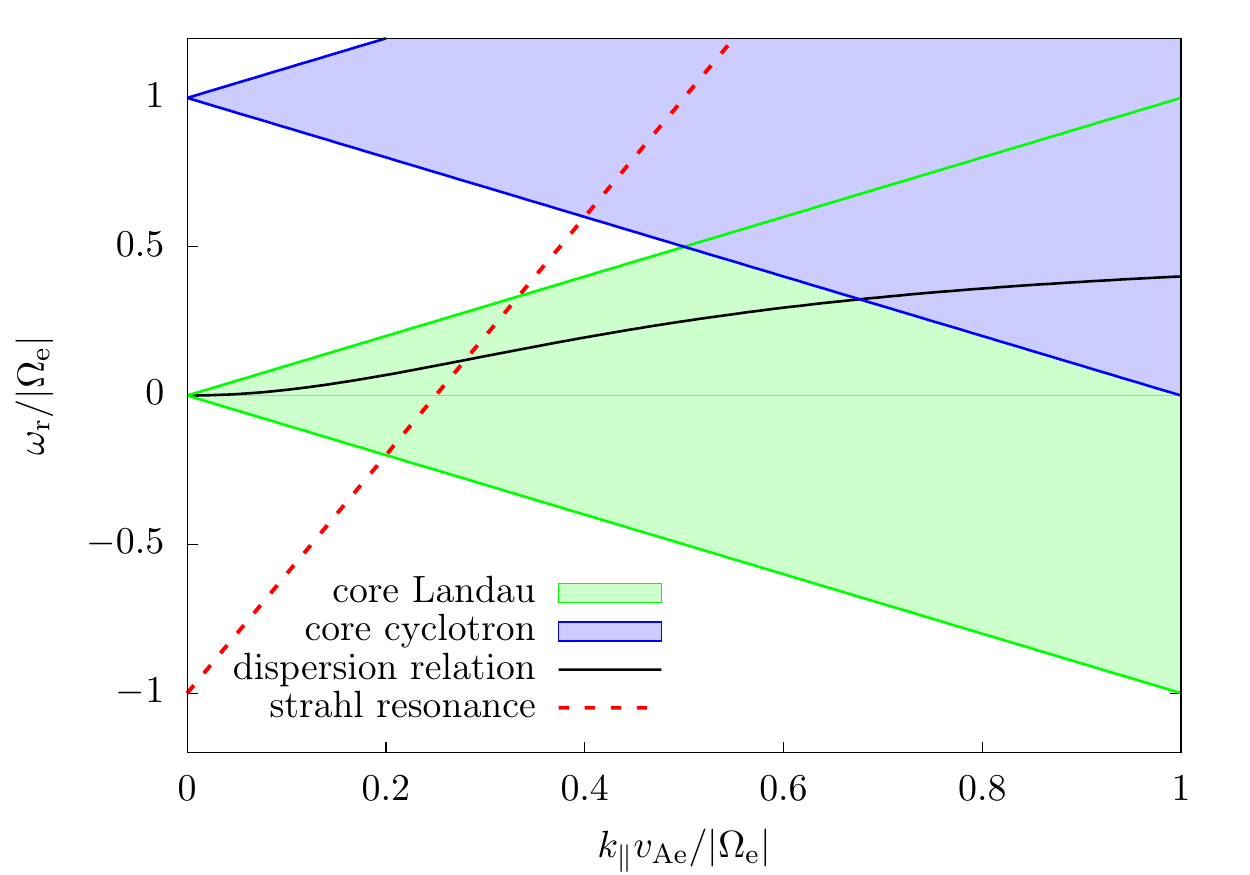}
\caption{Dispersion relation and resonance conditions for the FM/W mode with $\theta=60^{\circ}$ in the $\beta_{\mathrm c}\sim 1$ case. The black line shows Equation~(\ref{approxDR}). The blue and green areas show Equations~(\ref{corecond}) and (\ref{coreLcond}), respectively, and the red line shows Equation~(\ref{strahlcond}) with $U_{0\mathrm s}=4w_{\mathrm c}$. We use $w_{\mathrm c}=v_{\mathrm{Ae}}$.}
\label{resonances2}
\end{figure}
For  $\beta_{\mathrm c}\sim 1$, the dispersion relation lies below the Landau-damping threshold from $k_{\parallel}=0$ to $k_{\parallel}=k_{\mathrm{crit}}$. This situation is illustrated in Figure~\ref{resonances2}.
We now determine the instability threshold in this regime by balancing the destabilizing effects of the strahl against the stabilizing effects of core Landau damping. 
Under these assumptions, the FM/W mode is unstable if there is a wavenumber range in which 
\begin{equation}\label{generalcond}
\gamma_{k}^{\mathrm s,n=+1}+\gamma_k^{\mathrm c,n=0}>0.
\end{equation}
Using Equation~(\ref{growthrate}), we derive the instability criterion in Appendix~\ref{app_threshold}. We find that the FM/W mode is unstable if
\begin{equation}\label{cond1}
U_{0\mathrm s}\gtrsim\left[2\frac{n_{0\mathrm c}}{n_{0\mathrm s}}\sqrt{\frac{T_{0\mathrm s}}{T_{0\mathrm c}}} v_{\mathrm{Ae}}^2w_{\mathrm c}^2\frac{\left(1+\cos\theta\right)}{\left(1-\cos\theta \right)\cos\theta}\right]^{1/4}
\end{equation}
for the $\beta_{\mathrm c}\sim 1$ case. The minimum of the right-hand side of Equation~(\ref{cond1}) suggests that the lowest threshold occurs for $\cos\theta=-1+\sqrt{2}$, i.e., for $\theta\approx 65.6^{\circ}$.
The transition between the low-$\beta_{\mathrm c}$ case and the $\beta_{\mathrm c}\sim 1$ case occurs when $w_{\mathrm c}$ is large enough that the inequality in Equation~(\ref{coreLcond}) encompasses the entire plot of the dispersion relation. By combining Equation~(\ref{approxDR}) for $\theta=60^{\circ}$ and Equation~(\ref{coreLcond}), we find that this transition occurs when
\begin{equation}\label{transi}
w_{\mathrm c}\gtrsim \frac{v_{\mathrm{Ae}}}{2}.
\end{equation}

\begin{figure}
\epsscale{1.2}
\plotone{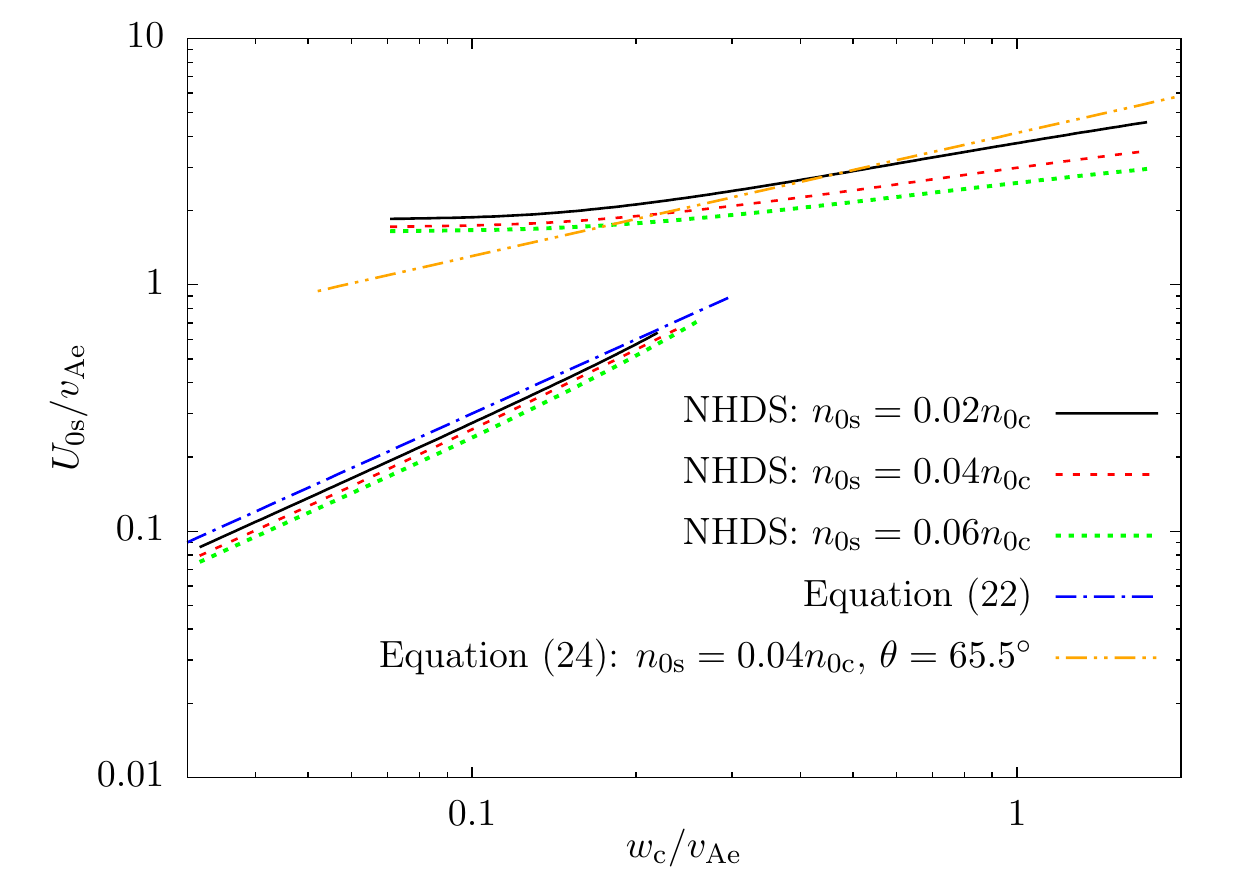}
\caption{Comparison of Equations~(\ref{cond2}) and (\ref{cond1}) with numerical solutions of the hot-plasma dispersion relation from our NHDS code. The blue and orange lines show Equations~(\ref{cond2}) and (\ref{cond1}), except that the $\gtrsim$ signs have been replaced with equal signs. We use $w_{\mathrm s}=2w_{\mathrm c}$ and $T_{0\mathrm p}=T_{0\mathrm c}$. For the numerical solutions, we show isocontours of constant maximum growth. The analytical solutions use $\theta=65.5^{\circ}$, while the numerical solutions are evaluated at the angle for which the lowest $U_{0\mathrm s}$ leads to a maximum growth rate of $\gamma_{\mathrm m}=10^{-3}|\Omega_{\mathrm e}|$.}
\label{eq34}
\end{figure}

We compare our analytical thresholds from Equations~(\ref{cond2}) and (\ref{cond1}) with numerical solutions of the hot-plasma dispersion relation from our NHDS code in Figure~\ref{eq34}.
Our analytical instability thresholds agree well with our numerical solutions. The transition between the low-$\beta_{\mathrm c}$ case and the $\beta_{\mathrm c}\sim 1$ case occurs at $w_{\mathrm c}\approx 0.2 v_{\mathrm{Ae}}$, which is slightly below our analytical finding in Equation~(\ref{transi}). We attribute this difference to inaccuracies based on our assumption of a discrete onset of Landau damping as soon as $\omega_{k\mathrm r}\le k_{\parallel}w_{\mathrm c}$.

\section{Comparison with Observations}\label{sect:observations}

\begin{figure}
\epsscale{1.2}
\plotone{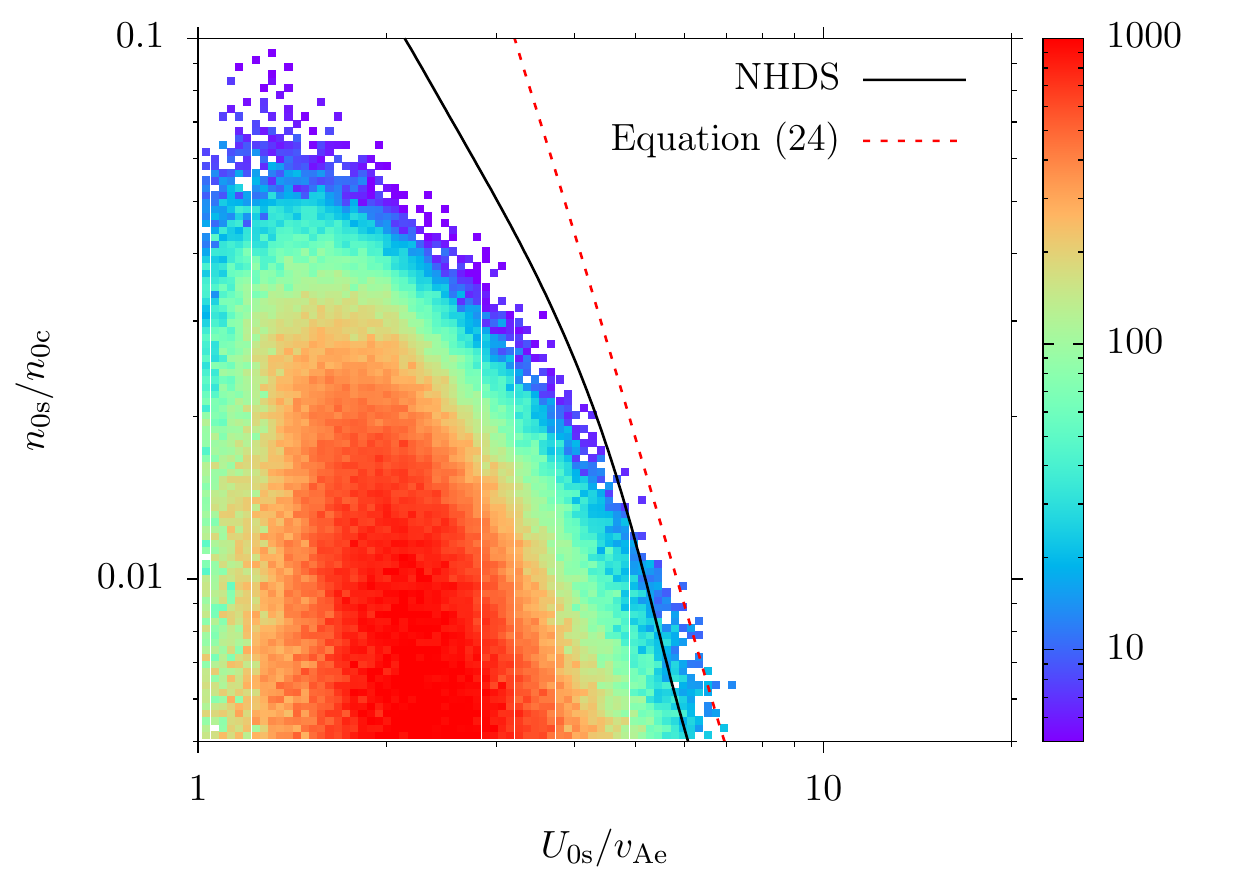}
\caption{Data distribution of the analyzed solar-wind interval in the $n_{0\mathrm s}/n_{0\mathrm c}$ vs. $U_{0\mathrm s}/v_{\mathrm{Ae}}$ plane. The color-coding shows the probability density (i.e., the number of counts per bin, normalized by the corresponding bin area to account for the logarithmic scaling of both plot axes) in arbitrary units. We determine the observed values for $n_{0\mathrm s}$ and $U_{0\mathrm s}$ by taking moments of the strahl distribution. The black line shows the isocontour of maximum growth rate $\gamma_{\mathrm m}=10^{-3}|\Omega_{\mathrm e}|$ for the oblique FM/W instability from our NHDS solutions. The red dashed line shows Equation~(\ref{cond1}) for  $\theta=65.5^{\circ}$, and $w_{\mathrm c}=v_{\mathrm{Ae}}=w_{\mathrm s}$. }
\label{fig_inst_thresholds}
\end{figure}

We use data from the 3DP instrument on board the \emph{Wind} spacecraft \citep{lin95}, obtained between 1995 and 1998.  With an automated routine \citep{pulupa14}, we fit the distribution function as a combination of a bi-Maxwellian distribution (core) and a bi-$\kappa$-distribution (halo):
\begin{equation}
f_{0\mathrm e}=f_{0\mathrm c}+f_{0\mathrm h}
\end{equation}
with
\begin{equation}
f_{0\mathrm c}=\frac{n_{0\mathrm c}}{\pi^{3/2}w_{\perp\mathrm c}^2w_{\parallel\mathrm c}}\exp\left(-\frac{(v_{\perp}-U_{0\perp\mathrm c})^2}{w_{\perp\mathrm c}^2}-\frac{(v_{\parallel}-U_{0\parallel\mathrm c})^2}{w_{\parallel\mathrm c}^2}\right)
\end{equation}
and
\begin{multline}
f_{0\mathrm h}=\frac{n_{0\mathrm h}}{w_{\perp\mathrm h}^2w_{\parallel\mathrm h}}\left[\frac{2}{\pi(2\kappa-3)}\right]^{3/2}\frac{\Gamma(\kappa+1)}{\Gamma(\kappa-1/2)}\\
\times \left\{1+\frac{2}{2\kappa-3}\left[\frac{(v_{\perp}-U_{0\perp\mathrm h})^2}{w_{\perp\mathrm h}^2}+\frac{(v_{\parallel}-U_{0\parallel\mathrm h})^2}{w_{\parallel\mathrm h}^2}\right]\right\}^{-(\kappa+1)},
\end{multline}
where $\Gamma(x)$ is the $\Gamma$-function, and the fit parameters are $n_{0\mathrm c}$, $w_{\perp\mathrm c}$, $w_{\parallel\mathrm c}$, $U_{0\perp\mathrm c}$, $U_{0\parallel\mathrm c}$, $n_{0\mathrm h}$, $w_{\perp\mathrm h}$, $w_{\parallel\mathrm h}$, $U_{0\perp\mathrm h}$, $U_{0\parallel\mathrm h}$, and $\kappa$. 
We determine the strahl bulk parameters as the result from subtracting the observed total electron distribution from the fit result and calculating the numerical moments of the remaining strahl distribution. In this way, we obtain the densities, relative drift speeds, temperatures, and temperature anisotropies of all electron species as well as the $\kappa$-index of the halo distribution.
We bin the data distribution in the $n_{0\mathrm s}/n_{0\mathrm c}$ vs. $U_{0\mathrm s}/v_{\mathrm {Ae}}$ plane and count the number of data points in each bin. We show the result in Figure~\ref{fig_inst_thresholds}. We limit ourselves to cases in which $n_{0\mathrm s}/n_{0\mathrm c}\ge 0.005$ since our automated method for the determination of the observed strahl bulk parameters suffers a loss of accuracy for smaller relative strahl densities.

We overplot numerical results for the instability threshold of the oblique FM/W instability for a maximum growth rate of $\gamma_{\mathrm m}=10^{-3}|\Omega_{\mathrm e}|$ from NHDS. We evaluate the threshold at the angle of propagation that leads to the maximum growth rate $\gamma_{\mathrm m}$. This angle varies between $51^{\circ}$ and $67^{\circ}$ in the shown range. We use the following free parameters: $\beta_{\mathrm p}=1$, $T_{0\mathrm c}=T_{0\mathrm p}$, $T_{0\mathrm s}=T_{0\mathrm p}$, and $v_{\mathrm{Ap}}/c=10^{-4}$. All species are isotropic. These choices represent typical values for these parameters at 1\,au consistent with our data set. In addition, we overplot Equation~(\ref{cond1}) for  $\theta=65.5^{\circ}$, and $w_{\mathrm c}=v_{\mathrm{Ae}}=w_{\mathrm s}$. 
We consider the parameter space to the lower left of the plotted instability threshold as the stable parameter space, while we consider the parameter space to the upper right of this curve as the unstable parameter space.
The instability threshold restricts the data in this parameter space to stable values, while only an insignificant number of data points populate the unstable parameter space. This finding is broadly consistent with our argument that the oblique FM/W instability sets the upper limit to $U_{0\mathrm s}$ in the solar wind. We also note that the numerical solution and our analytical solution agree reasonably well, especially at small $n_{0\mathrm s}/n_{0\mathrm c}$, the regime in which the strahl effect on the dispersion relation is negligible as assumed in our derivation. We note that some of the parameter combinations shown in Figure~\ref{fig_inst_thresholds} exhibit a bump-on-tail configuration when the model Maxwellian core and strahl distributions are summed, which can be unstable to other instabilities (see Section~\ref{elstat}). However, we neglect bump-on-tail instabilities in this paper, because, in a more realistic model, the total electron distribution function would be a monotonically decreasing function of $|v_{\parallel}|$. Under realistic solar-wind conditions, the number of halo electrons that are in resonance with the unstable FM/W waves is small. Therefore, we conjecture that our simple core/strahl model of the electron distribution provides a reasonable approximation of the oblique FM/W instability threshold. However, future investigations based on more realistic core/halo/strahl electron distributions will be needed in order to test this conjecture.

\section{Relation to Other Electron-driven Instabilities}\label{sect:others}

In this section, we discuss the relevance of other electron-driven instabilities to the evolution of the electron strahl in the solar wind. We specifically address the consistency of these alternative instabilities with the strahl-scattering scenario described in Section~\ref{sect:scenario}. Our reasoning relies on the considerations presented in Section~\ref{sect:framework} of the diffusion paths with respect to the phase speed as well as a careful analysis with NHDS.

\subsection{Whistler Heat-flux Instability}

\begin{figure}
\epsscale{1.1}
\plotone{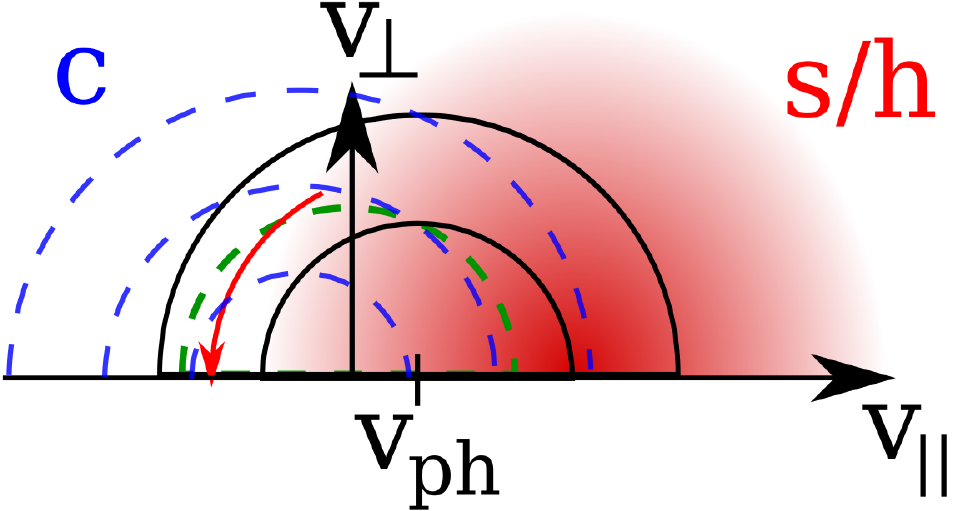}
\caption{Diffusion paths for the parallel whistler heat-flux instability. The electron beam must fulfill $U_{0\mathrm s}>v_{\mathrm{ph}}$ and be hot enough to have a sufficient number of electrons at $v_{\parallel}<0$. Typically, the strahl does not  fulfill this latter requirement, while the halo can be hot enough. The core distribution is shown as blue dashed semicircles and the strahl or halo distribution as a red shaded semicircle. The particle diffusion path (red arrow) is locally circular (indicated by black semicircles) about the parallel phase speed $v_{\mathrm{ph}}=\omega_{k\mathrm r}/k_{\parallel}$. The green circle segment indicates constant kinetic energy. Particles at $v_{\parallel}<0$ diffuse and lose kinetic energy for this instability.}
\label{fig_diff_paths_whistler}
\end{figure}

The parallel-propagating FM/W wave is purely right-handed in polarization (i.e., $E_{k,\mathrm l}=E_{kz}=0$). Equation~(\ref{eq:psink}) for $k_{\perp}=0$, therefore, requires that the only contributing resonant interaction is the cyclotron resonance with $n=-1$.
The resonant particles driving this instability must move in the opposite direction along the magnetic field as the wave (i.e., $v_{\parallel}<0$ in our convention) in order to fulfill the resonance condition, Equation~(\ref{rescond}). This property characterizes the halo rather than the strahl. If unstable, this mode corresponds to the whistler heat-flux instability \citep{gary77,gary99,wilson09,wilson13,shaaban18}. The diffusion paths for this instability are shown in Figure~\ref{fig_diff_paths_whistler} \citep[see also][]{shaaban19}.
The resonant particles diffuse toward smaller values of $v_{\perp}$ and form a tail-like structure in the distribution. This behavior does not agree with the scenario that strahl scattering forms the halo distribution function as discussed in Section~\ref{sect:scenario}. For these reasons, we exclude the parallel whistler heat-flux instability as a candidate for a plasma instability that scatters the strahl into the halo. This instability can, however, be relevant for the regulation of the halo heat flux in the solar wind. Its threshold then also depends on the halo anisotropy \citep{shaaban18b}. We note that a reduction of $U_{0\mathrm c}$ can lead to an indirect reduction of $U_{0\mathrm s}$ by the fulfillment of quasi-neutrality according to Equation~(\ref{zerocurrents}). This indirect effect is also relevant for the ion-acoustic heat-flux instability and for the KAW heat-flux instability. Evidence for the whistler heat-flux instability was found in measurements of solar-wind core and halo electrons \citep{tong19}.


\subsection{Lower-hybrid Fan Instability}

\begin{figure}
\epsscale{1.1}
\plotone{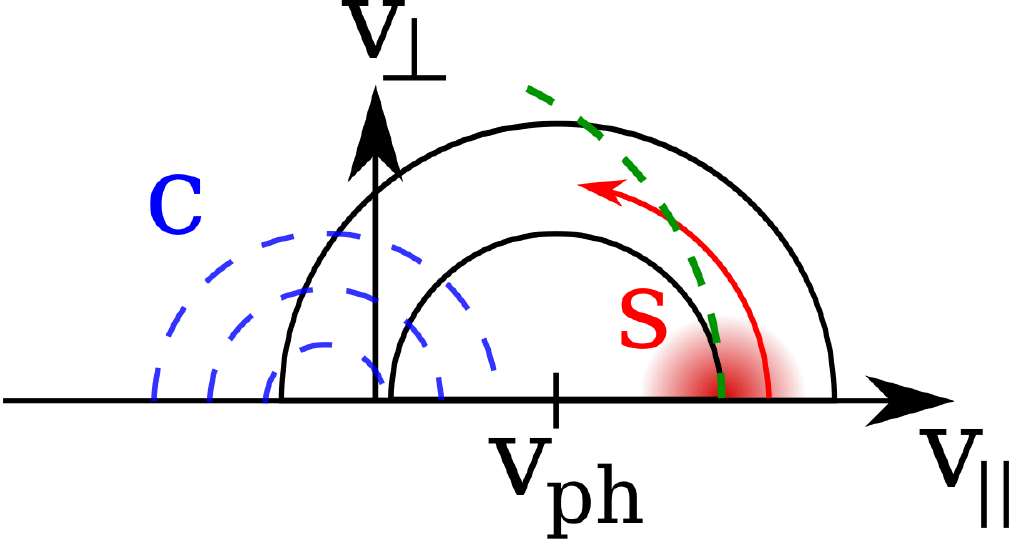}
\caption{Diffusion paths in a core--strahl electron distribution function for the lower-hybrid fan instability. The core distribution is shown as blue dashed semicircles and the strahl distribution as a red shaded semicircle. The particle diffusion path (red arrow) is locally circular (indicated by black semicircles) about the parallel phase speed $v_{\mathrm{ph}}=\omega_{k\mathrm r}/k_{\parallel}$. The green circle segment indicates constant kinetic energy. $\beta_{\mathrm c}$ is small so that the distribution functions of core and strahl do not overlap sufficiently for strong core Landau damping to suppress the instability.}
\label{fig_diff_paths_fan}
\end{figure}

In the limit $\cos^2\theta <m_{\mathrm e}/m_{\mathrm p}$, the FM/W-mode branch corresponds to the lower-hybrid mode as shown in Section~\ref{sect:criterion}. 
The fan instability of the lower-hybrid mode is driven by the $n=+1$ resonance of strahl electrons like the oblique FM/W instability \citep{omelchenko94,shevchenko10}. It scatters particles about the parallel phase speed of the lower-hybrid wave, $v_{\mathrm{ph}}=\omega_{\mathrm{LH}}/k_{\parallel}$, and is, thus, capable of scattering  strahl electrons into the halo. The highly oblique lower-hybrid mode has a strong electrostatic component and relatively low frequencies compared to the moderate-$\theta$ FM/W mode. Therefore, it is prone to strong core Landau damping if $w_{\mathrm c}$ is large enough that the core provides a significant number of electrons at $v_{\parallel}\approx v_{\mathrm{ph}}$. In this case, the resonance speed for Landau-resonant core electrons lies deep within the core distribution function. With increasing $w_{\mathrm c}$, the phase speed of the lower-hybrid wave increases slightly due to thermal corrections to its dispersion relation. This effect can overcompensate the increasing number of Landau-resonant core electrons at the strahl resonance speed. However, the growth rate of the highly oblique lower-hybrid mode is still less than the growth rate of the oblique FM/W instability by about two orders of magnitude under typical solar-wind conditions. As $U_{0\mathrm s}$ increases, $|U_{0\mathrm c}|$ must increase so that the parallel current vanishes as per Equation~(\ref{zerocurrents}).
 In cases with very low $\beta_{\mathrm c}$, the lower-hybrid fan instability is a good candidate for self-induced strahl scattering. This situation is illustrated in Figure~\ref{fig_diff_paths_fan}.
The instability is related to the oblique FM/W instability since it is driven by the same resonance effect and transitions into the oblique FM/W mode for smaller $\theta$. We note that some authors use a broader definition of the term \emph{fan instabilities} for all instabilities driven by an $n=+1$ resonance \citep[e.g., ][]{krafft03}. 

\subsection{Ion-acoustic Heat-flux Instability and Kinetic-Alfv\'en-wave Heat-flux Instability}

\begin{figure}
\epsscale{1.1}
\plotone{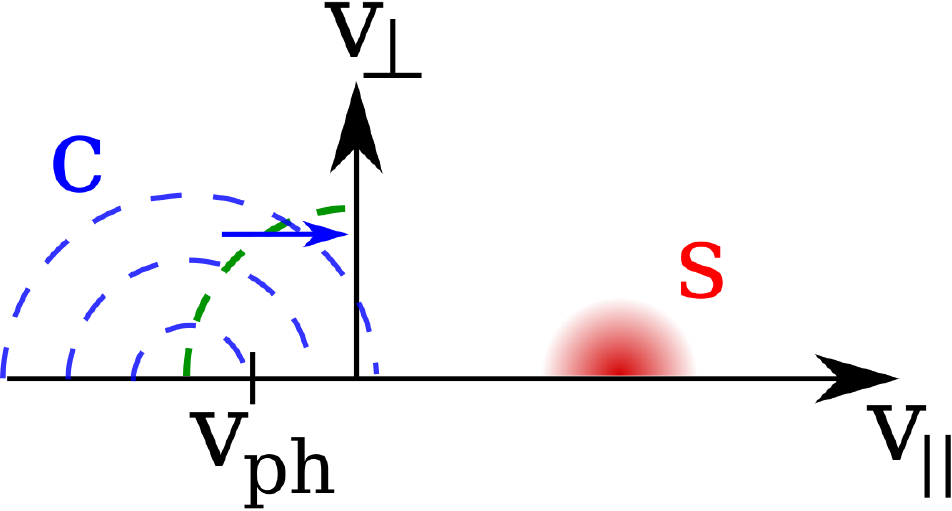}
\caption{Diffusion paths in a core--strahl electron distribution function for the ion-acoustic heat-flux instability and for the KAW heat-flux instability. The core distribution is shown as blue dashed semicircles and the strahl distribution as a red shaded semicircle. The particle diffusion path (blue arrow) is purely parallel to the magnetic field and occurs at $v_{\parallel}=v_{\mathrm{ph}}$.  In the case of the ion-acoustic heat-flux instability, $T_{0\mathrm c}\gg T_{0\mathrm p}$ so that the phase speed is large enough to overcome proton Landau damping. For the KAW heat-flux instability, the core temperature is typically small $\beta_{\mathrm c}\leq 10^{-3}$.}
\label{fig_diff_paths_iaw}
\end{figure}

The ion-acoustic heat-flux instability \citep{gary78} has a comparable instability threshold to the threshold of the oblique FM/W instability under certain conditions. However, this instability acts on particles with $v_{\parallel}\approx v_{\mathrm{ph}}$. The parallel phase speed of the ion-acoustic mode is much less than both $w_{\mathrm c}$ and the typical strahl speed in the solar wind. Therefore, it is more likely that this instability is driven by the core drift, which is directed in the opposite direction of the strahl speed according to Equation~(\ref{zerocurrents}). For this case, we illustrate the quasilinear diffusion of particles under the action of the ion-acoustic instability in Figure~\ref{fig_diff_paths_iaw}.
The ion-acoustic heat-flux instability is driven by the Landau resonance of the core electrons, which leads to a diffusion of core electrons at $v_{\parallel}=v_{\mathrm{ph}}$ toward smaller $|v_{\parallel}|$. This instability is not a candidate to explain strahl scattering into the halo since it does not increase $v_{\perp}$ of strahl electrons. We also note that large $T_{0\mathrm c}\gg T_{0\mathrm p}$ are required in order to increase  $v_{\mathrm{ph}}$ of the ion-acoustic mode and to avoid proton Landau damping, which would otherwise efficiently suppress this instability. The maximum growth rate of the ion-acoustic heat-flux instability occurs in parallel propagation.

The same instability mechanism is responsible for the kinetic-Alfv\'en-wave (KAW) heat-flux instability \citep{gary75}. In the parameter range explored by \citet{gary75}, the KAW heat-flux instability requires low $\beta_{\mathrm c} \lesssim 10^{-3}$ to have the lowest threshold of all electron-drift driven instabilities. It propagates at very large angles $\theta$ with respect to $\vec B_0$. The nonzero $E_{kz}$ of the KAW allows for the electron core to drive the mode unstable through the Landau resonance as in the case of the ion-acoustic heat-flux instability. Likewise, this instability does not scatter strahl particles into the halo.

\subsection{Electrostatic Electron-beam Instability}\label{elstat}

\begin{figure}
\epsscale{1.1}
\plotone{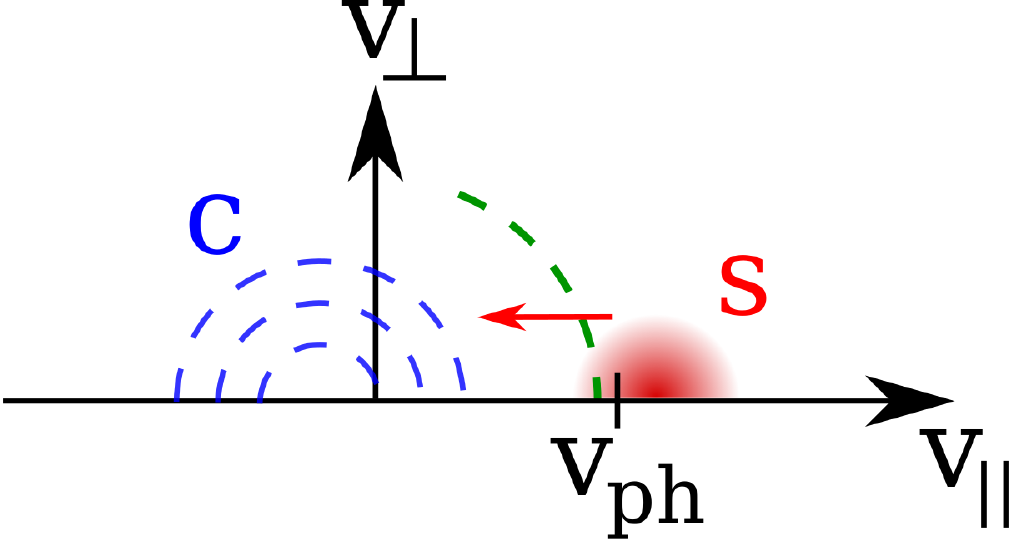}
\caption{Diffusion paths in a core--strahl electron distribution function for the electrostatic electron-beam instability. The core distribution is shown as blue dashed semicircles and the strahl distribution as a red shaded semicircle. The particle diffusion path (red arrow) is purely parallel to the magnetic field and occurs at $v_{\parallel}=v_{\mathrm{ph}}$. The green circle segment indicates constant kinetic energy. $\beta_{\mathrm c}$ is small so that core-Landau damping is inefficient in suppressing the instability. The strahl speed has to be slightly greater than $v_{\mathrm{ph}}$ in order to yield an kinetic-energy loss of strahl electrons during the diffusion.}
\label{fig_diff_paths_elstat}
\end{figure}

The electrostatic electron-beam instability \citep{gary78} is a low-$\beta_{\mathrm c}$ instability that propagates into the direction of the strahl and has maximum growth  parallel to $\vec B_0$. Its typical phase speed is  $v_{\mathrm{ph}}\lesssim U_{0\mathrm s}$. For $\beta_{\mathrm c}\lesssim 0.1$, it has lower thresholds than the other instabilities, as long as the strahl forms a bump-on-tail configuration rather than a shoulder of the distribution while all other parameters are kept at representative solar-wind values. It is driven by the Landau resonance of the strahl electrons. We illustrate this instability in Figure~\ref{fig_diff_paths_elstat}.
The Landau-resonant diffusion of strahl electrons leads to an excitation of this instability only if the diffusion paths down the gradients of the strahl distribution function are directed to lower kinetic energies. For this reason, $U_{0\mathrm s}$ has to be slightly greater than $v_{\mathrm{ph}}$ as illustrated in Figure~\ref{fig_diff_paths_elstat}.
This instability only reduces $v_{\parallel}$ and does not increase $v_{\perp}$ of the resonant strahl electrons and is thus not a candidate mechanism to scatter strahl electrons into the halo. We also note that, if this instability were to operate in the solar wind, it would lead to a quasilinear flattening of the electron distribution function in at least some narrow range of parallel velocities near the strahl velocity. However, the measured electron distribution functions are strongly decreasing functions of $v_{\parallel}$ near the strahl velocity \citep{pilipp87,marsch06}.

\section{Conclusions}\label{sec:conc}

Electron distribution functions in the solar wind consist of three components: a core, halo, and  strahl. The relative drifts among these populations carry a significant heat flux from the solar corona into the heliosphere. It is, therefore, of great importance for global solar-wind models to understand  the regulation of the relative drifts between these electron components. Observations indicate that strahl electrons are continuously transferred into the halo. We propose a mechanism that explains this scattering as the consequence of a self-induced excitation of the oblique FM/W instability. In this scenario, we assume that electrons are accelerated to high energies in the solar corona. We furthermore assume that the conservation of the magnetic moment in the widening magnetic-field structure of a coronal hole focuses these energetic electrons into the antisunward direction, forming the electron strahl. Based on these assumptions, we find that the strahl itself then quasi-continuously excites an instability of the oblique FM/W wave that scatters strahl electrons into the halo and generates plasma waves with wavelengths between the ion and electron kinetic scales.
This instability reduces the strahl density, increases the halo density, and limits the strahl heat flux. 

In Section~\ref{sect:fmw}, we derive analytical expressions for the thresholds of the oblique FM/W instability in both the low-$\beta_{\mathrm c}$ regime and the $\beta_{\mathrm c}\sim 1$ regime. In the low-$\beta_{\mathrm c}$ regime, the strahl excites FM/W waves when $U_{0\mathrm s}$ is large enough that the strahl can resonate with the waves at wavenumbers and frequencies at which core Landau damping and core cyclotron damping are negligible. At $\beta_{\mathrm c}\sim 1$, on the other hand, core Landau damping of FM/W waves cannot be avoided, and cyclotron driving by the strahl must overcome this core Landau damping in order to make the FM/W waves unstable.

In Section~\ref{sect:observations}, we compare the instability thresholds of the oblique FM/W instability with direct in-situ measurements from the \emph{Wind} spacecraft.  We find that the instability limits the data distribution to the stable regime. This finding corroborates our hypothesis that the oblique FM/W instability indeed limits the strahl speed and heat flux in the solar wind. In the future, we will study the instability of observed electron distributions with our ALPS code \citep{verscharen18} without relying on the assumption of a Maxwellian shape of the electron components' distribution in Equation~(\ref{Maxwell}), which is also made in our NHDS solutions.

Other electron-driven instabilities (whistler heat-flux, lower-hybrid fan, ion-acoustic heat-flux, kinetic-Alfv\'en-wave heat-flux, and electrostatic electron-beam instabilities) are either not capable of scattering strahl electrons into the halo or (in the case of the lower-hybrid fan instability) have growth rates much smaller than the oblique FM/W instability.

Fully kinetic simulations, such as Vlasov or particle-in-cell simulations, and simulations of the quasilinear diffusion equations \citep[e.g., using the methods presented by][]{pongkitiwanichakul14} will allow us to model the nonlinear evolution of the strahl--halo system in a future analysis.  \emph{Parker Solar Probe (PSP)}  measures electron distribution functions and waves in the close vicinity of the Sun. We predict that \emph{PSP} will encounter the low-$\beta_{\mathrm c}$ regime of the oblique FM/W instability in which its threshold is given by Equation~(\ref{cond2}). In this case, we predict the presence of copious FM/W waves with $\theta\approx 60^{\circ}$ and $\omega_{\mathrm r}\approx 0.5|\Omega_{\mathrm e}|$. PSP may also encounter the point at which the strahl speed crosses the threshold of the FM/W instability for the first time. This discovery will help us understand the range of distances where the oblique FM/W instability is relevant. The \emph{Solar Orbiter} spacecraft will link the observed high-cadence and high-resolution in-situ electron properties with the associated source regions in the corona in order to improve our understanding of the global evolution of electrons in the solar wind.

\acknowledgements

We appreciate helpful discussions with Sofiane Bourouaine, Georgie Graham, Allan Macneil, Rob Wicks, and Chris Owen. This work was discussed at the 2019 ESAC Solar Wind Electron Workshop, which was supported by the Faculty of the European Space Astronomy Centre (ESAC). D.V.~is supported by the STFC Ernest Rutherford Fellowship ST/P003826/1 and STFC Consolidated Grant ST/S000240/1. This work is supported in part by NASA contract NNN06AA01C, NASA grant NNX16AG81G, NASA grant NNX17AI18G, NASA grant 80NSSC19K0829, and NSF SHINE grant 1460190. Work at UC Berkeley is supported in part by NASA grant NNX14AC09G, NASA grant NNX16AI59G, and NSF SHINE grant 1622498. S.D.B.~acknowledges support from the Leverhulme Trust Visiting Professorship program.

\appendix

\section{Analytical Instability Criterion for $\beta_{\mathrm c}\sim 1$}\label{app_threshold}

Our calculation of the instability threshold is based on Equation~(\ref{total_growthrate}) for the case in which strahl driving balances with the stabilizing effects of core Landau damping. We first calculate expressions for $|\psi_k^{\mathrm c,n=0}|^2/W_k$ and $|\psi_k^{\mathrm s,n=+1}|^2/W_k$.  We then use these expressions to derive $\gamma_k^{\mathrm c,n=0}$ and $\gamma_k^{\mathrm s,n=+1}$. In order to simplify Equation~(\ref{total_growthrate}), we approximate $\omega_{k\mathrm r}$ using the cold-plasma dispersion relation for an electron--proton plasma \citep{stix92}, recognizing that this will introduce some error into our results.  
 
\subsection{Polarization of Oblique FM/W Waves}

In the cold-plasma dispersion relation \citep{stix92},
\begin{equation}\label{iEx}
\frac{iE_{kx}}{E_{ky}}=\frac{n^2-S}{D}
\end{equation}
and
\begin{equation}\label{iEz}
\frac{iE_{kz}}{E_{ky}}=\frac{n^4\cos^2\theta-n^2S\left(1+\cos^2\theta\right)+S^2-D^2}{Dn^2\cos\theta\sin\theta},
\end{equation}
where $n\equiv kc/\omega_k$ is the refractive index and, in the whistler-wave regime,
\begin{equation}
S\simeq \frac{\omega_{\mathrm{pe}}^2}{\Omega_{\mathrm e}^2-\omega_k^2}
\end{equation}
and
\begin{equation}
D\simeq -S\frac{\Omega_{\mathrm e}}{\omega_k}.
\end{equation}
Using Equation~(\ref{approxDR}), Equations~(\ref{iEx}) and (\ref{iEz}) simplify to
\begin{equation}\label{Ekx}
\frac{iE_{kx}}{E_{ky}}\simeq \frac{1+k^2d_{\mathrm e}^2\sin^2\theta}{\cos\theta}
\end{equation}
and
\begin{equation}\label{Ekz}
\frac{iE_{kz}}{E_{ky}}\simeq k^2d_{\mathrm e}^2\sin\theta.
\end{equation}
We choose the coordinate system in which $\phi=0$ so that $k_{\perp}=k_x$ and $k_y=0$.
With the use of Equations~(\ref{eq:psink}), (\ref{approxDR}), (\ref{Ekx}), and (\ref{Ekz}), we find
\begin{equation}
\frac{\left|\psi_k^{\mathrm c,n=0}\right|^2}{W_k}\simeq \left[\frac{v_{\parallel}}{v_{\perp}}k^2d_{\mathrm e}^2\sin\theta\, J_0\left(x_{\mathrm c}\right)+J_1\left(x_{\mathrm c}\right)\right]^2\frac{\left|E_{ky} \right|^2}{W_k}.
\end{equation}
Since $iE_{kx}/E_{ky} \gg iE_{kz}/E_{ky}$ and $k_{\perp}\rho_{\mathrm e}\ll1$, we retain only the term proportional to $J_0$ in $\psi_k^{\mathrm s,n=+1}$. With the use of  Equations~(\ref{eq:psink}), (\ref{approxDR}), (\ref{Ekx}), and (\ref{Ekz}), we then obtain 
\begin{equation}
\frac{\left|\psi_k^{\mathrm s,n=+1}\right|^2}{W_k}\simeq \frac{1}{4}\frac{\left(1-\cos\theta+k^2d_{\mathrm e}^2\sin^2\theta\right)^2}{\cos^2\theta}J_0^2\left(x_{\mathrm s}\right)\frac{\left|E_{ky} \right|^2}{W_k}.
\end{equation}

\subsection{Dispersion Relation and Resonance Condition}

By combining Equations~(\ref{strahlcond}) and (\ref{approxDR}), we find for the resonance condition
\begin{equation}\label{rescondmod}
\frac{k^2d_{\mathrm e}^2\cos \theta}{1+k^2d_{\mathrm e}^2}-kd_{\mathrm e}\cos\theta \frac{U_{0\mathrm s}}{v_{\mathrm{Ae}}}+1=0.
\end{equation}
We make the simplifying approximation that 
\begin{equation}\label{epsilon}
\epsilon\equiv \frac{v_{\mathrm{Ae}}}{U_{0\mathrm s}}\ll 1.
\end{equation}
Solving Equation~(\ref{rescondmod}) using the method of dominant balance \citep{bender99}, we obtain
\begin{equation}\label{reskpar}
k_{\parallel}d_{\mathrm e}\approx \frac{v_{\mathrm{Ae}}}{U_{0\mathrm s}}\left(1+\frac{v_{\mathrm{Ae}}^2}{U_{0\mathrm s}^2}\frac{1}{\cos\theta}+\dots\right).
\end{equation}
Upon substituting Equation~(\ref{reskpar}) into Equation~(\ref{approxDR}), we find that the parallel phase velocity of the resonant wave is 
\begin{equation}\label{phasespeed}
\frac{\omega_{k\mathrm r}}{k_{\parallel}}\approx \frac{v_{\mathrm{Ae}}^2}{U_{0\mathrm s}\cos\theta}\left[1+\frac{v_{\mathrm{Ae}}^2}{U_{0\mathrm s}^2\cos^2\theta}\left(\cos\theta-1\right)+\dots\right]
\end{equation}
at the resonant wavenumber and frequency.

\subsection{Evaluation of Growth and Damping Rates}

We simplify the integrals in Equation~(\ref{growthrate}) by exploiting the $\delta$-function and  the Bessel-function identities\footnote{Our result in Equations~(\ref{gam11}) and (\ref{gam12}) can also be obtained by first approximating the $J_{\nu}(x)$ term using their small-$x$ expansions, which avoids the use of the Bessel-function identities in Equations~(\ref{Bessel1}) through (\ref{Bessel3}).}
\begin{equation}\label{Bessel1}
\int\limits_0^{\infty}\mathrm dv_{\perp}\,v_{\perp}J_{\nu}^2\left(\frac{k_{\perp}v_{\perp}}{\Omega_j}\right)e^{-v_{\perp}^2/w_j^2}=\frac{w_j^2}{2}e^{-\lambda_j}I_{\nu}(\lambda_j),
\end{equation}
\begin{multline}
\int\limits_0^{\infty}\mathrm dv_{\perp}\,v_{\perp}^3J_{\nu}^2\left(\frac{k_{\perp}v_{\perp}}{\Omega_j}\right)e^{-v_{\perp}^2/w_j^2}=\frac{w_j^4}{2}e^{-\lambda_j}\\
\times \left\{I_{\nu}(\lambda_j)-\lambda_j\left[I_{\nu}(\lambda_j)-I_{\nu}^{\prime}(\lambda_j)\right]\right\},
\end{multline}
and
\begin{multline}\label{Bessel3}
\int \limits _0^{\infty}\mathrm dv_{\perp} v_{\perp}^2J_0\left(\frac{k_{\perp}v_{\perp}}{\Omega_j}\right)J_1\left(\frac{k_{\perp}v_{\perp}}{\Omega_j}\right)e^{-v_{\perp}^2/w_j^2}=\frac{k_{\perp}w_j^4}{4\Omega_j}e^{-\lambda_j}\\
\times \left[I_1(\lambda_j)-I_0(\lambda_j)\right],
\end{multline}
where $I_{\nu}$ is the modified Bessel function of order $\nu$ and $\lambda_j\equiv k_{\perp}^2w_j^2/2\Omega_j^2$. Equations~(\ref{Bessel1})--(\ref{Bessel3}) follow from the identity \citep{watson22}
\begin{equation}
\int\limits_0^{\infty}\mathrm dt\,tJ_{\nu}(at)J_{\nu}(bt)e^{-p^2t^2}=\frac{1}{2p^2}\exp\left(-\frac{a^2+b^2}{4p^2}\right)I_{\nu}\left(\frac{ab}{2p^2}\right)
\end{equation}
in the way presented by \citet{stix92}, which is valid for $\mathrm{Re}(\nu)>-1$ and $|\mathrm{arg}\,p|<\pi/4$.  Given Equations~(\ref{epsilon}) and (\ref{reskpar}), $\lambda_{\mathrm c}=0.5 \beta_{\mathrm c} \epsilon^2\tan^2\theta+\dots$ We take $\tan\theta$ and $\beta_{\mathrm c}$ to be of order unity, and thus $\lambda_{\mathrm c}\ll1$. We take $w_{\mathrm s}\sim w_{\mathrm c}$, and thus $\lambda_{\mathrm s}$ is also $\ll 1$. With the use of Equations~(\ref{strahlcond}), (\ref{reskpar})--(\ref{Bessel3}), and our assumptions that $\omega_{k\mathrm r}>0$, $k_{\parallel}>0$, we find that to leading order in $\epsilon$,
\begin{equation}\label{gam11}
\frac{\gamma_k^{\mathrm c,n=0}}{\omega_{k\mathrm r}}\simeq - \left(\frac{\omega_{k\mathrm r}}{k_{\parallel}w_{\mathrm c}}\right)\left(\frac{\omega_{\mathrm{pc}}}{\omega_{k\mathrm r}}\right)^2\frac{\left|E_{ky} \right|^2/W_k}{8\sqrt{\pi}} \lambda_{\mathrm c}
\end{equation}
and
\begin{multline}\label{gam12}
\frac{\gamma_k^{\mathrm s,n=+1}}{\omega_{k\mathrm r}}\simeq  \left(\frac{\omega_{k\mathrm r}}{k_{\parallel}w_{\mathrm c}}\right)\left(\frac{\omega_{\mathrm{pc}}}{\omega_{k\mathrm r}}\right)^2\frac{\left|E_{ky} \right|^2/W_k}{8\sqrt{\pi}}\\
\times\frac{n_{0\mathrm s}}{n_{0\mathrm c}}\sqrt{\frac{T_{0\mathrm c}}{T_{0\mathrm s}}}\frac{\left|\Omega_{\mathrm e}\right|}{4\omega_{k\mathrm r}}\frac{\left(1-\cos\theta\right)^2}{\cos^2\theta}.
\end{multline}
Using these closed expressions for $\gamma_k^{\mathrm c,n=0}$ and $\gamma_k^{\mathrm s,n=+1}$, the instability condition in Equation~(\ref{generalcond}) then translates to
\begin{equation}
\frac{n_{0\mathrm s}}{n_{0\mathrm c}}\sqrt{\frac{T_{0\mathrm c}}{T_{0\mathrm s}}}\frac{U_{0\mathrm s}^2}{4v_{\mathrm{Ae}}^2}\frac{\left(1-\cos\theta\right)^2}{\cos\theta}\gtrsim \frac{w_{\mathrm c}^2}{2U_{0\mathrm s}^2}\tan^2\theta,
\end{equation}
which furthermore simplifies to 
\begin{equation}
\frac{U_{0\mathrm s}^4}{v_{\mathrm{Ae}}^4}
\gtrsim 2\frac{n_{0\mathrm c}}{n_{0\mathrm s}}\sqrt{\frac{T_{0\mathrm s}}{T_{0\mathrm c}}}\frac{w_{\mathrm c}^2}{v_{\mathrm{Ae}}^2}\frac{\left(1+\cos\theta\right)}{\left(1-\cos\theta \right)\cos\theta}.
\end{equation}
This criterion can be rewritten as Equation~(\ref{cond1}).

\bibliographystyle{apj}
\bibliography{electron_strahl_rev_rev}

\begin{thebibliography}{85}
\expandafter\ifx\csname natexlab\endcsname\relax\def\natexlab#1{#1}\fi

\bibitem[{{Alexandrova} {et~al.}(2012){Alexandrova}, {Lacombe}, {Mangeney},
  {Grappin}, \& {Maksimovic}}]{alexandrova12}
{Alexandrova}, O., {Lacombe}, C., {Mangeney}, A., {Grappin}, R., \&
  {Maksimovic}, M. 2012, \apj, 760, 121

\bibitem[{{Alexandrova} {et~al.}(2009){Alexandrova}, {Saur}, {Lacombe},
  {Mangeney}, {Mitchell}, {Schwartz}, \& {Robert}}]{alexandrova09}
{Alexandrova}, O., {Saur}, J., {Lacombe}, C., {et~al.} 2009, \prl, 103, 165003

\bibitem[{{Bale} {et~al.}(2013){Bale}, {Pulupa}, {Salem}, {Chen}, \&
  {Quataert}}]{bale13}
{Bale}, S.~D., {Pulupa}, M., {Salem}, C., {Chen}, C.~H.~K., \& {Quataert}, E.
  2013, \apjl, 769, L22

\bibitem[{{Bender} \& {Orszag}(1999)}]{bender99}
{Bender}, C.~M., \& {Orszag}, S.~A. 1999, {Advanced Mathematical Methods for
  Scientists and Engineers I} ({New York, USA}: {Springer})

\bibitem[{{Chandran} {et~al.}(2010){Chandran}, {Pongkitiwanichakul},
  {Isenberg}, {Lee}, {Markovskii}, {Hollweg}, \& {Vasquez}}]{chandran10b}
{Chandran}, B.~D.~G., {Pongkitiwanichakul}, P., {Isenberg}, P.~A., {et~al.}
  2010, \apj, 722, 710

\bibitem[{{Chen} {et~al.}(2012){Chen}, {Salem}, {Bonnell}, {Mozer}, \&
  {Bale}}]{chen12}
{Chen}, C.~H.~K., {Salem}, C.~S., {Bonnell}, J.~W., {Mozer}, F.~S., \& {Bale},
  S.~D. 2012, \prl, 109, 035001

\bibitem[{{Detering} {et~al.}(2005){Detering}, {Rozmus}, {Brantov},
  {Bychenkov}, {Capjack}, \& {Sydora}}]{detering05}
{Detering}, F., {Rozmus}, W., {Brantov}, A., {et~al.} 2005, Phys.~Plasmas, 12,
  012321

\bibitem[{{Feldman} {et~al.}(1976){Feldman}, {Asbridge}, {Bame}, {Gary}, \&
  {Montgomery}}]{feldman76}
{Feldman}, W.~C., {Asbridge}, J.~R., {Bame}, S.~J., {Gary}, S.~P., \&
  {Montgomery}, M.~D. 1976, \jgr, 81, 2377

\bibitem[{{Feldman} {et~al.}(1975){Feldman}, {Asbridge}, {Bame}, {Montgomery},
  \& {Gary}}]{feldman75}
{Feldman}, W.~C., {Asbridge}, J.~R., {Bame}, S.~J., {Montgomery}, M.~D., \&
  {Gary}, S.~P. 1975, \jgr, 80, 4181

\bibitem[{{Fitzenreiter} {et~al.}(1998){Fitzenreiter}, {Ogilvie}, {Chornay}, \&
  {Keller}}]{fitzenreiter98}
{Fitzenreiter}, R.~J., {Ogilvie}, K.~W., {Chornay}, D.~J., \& {Keller}, J.
  1998, \grl, 25, 249

\bibitem[{{Gary}(1978)}]{gary78}
{Gary}, S.~P. 1978, J.~Plasma Phys., 20, 47

\bibitem[{{Gary}(1979)}]{gary79}
---. 1979, J.~Plasma Phys., 21, 361

\bibitem[{{Gary} \& {Feldman}(1977)}]{gary77}
{Gary}, S.~P., \& {Feldman}, W.~C. 1977, \jgr, 82, 1087

\bibitem[{{Gary} {et~al.}(1975{\natexlab{a}}){Gary}, {Feldman}, {Forslund}, \&
  {Montgomery}}]{gary75}
{Gary}, S.~P., {Feldman}, W.~C., {Forslund}, D.~W., \& {Montgomery}, M.~D.
  1975{\natexlab{a}}, \grl, 2, 79

\bibitem[{{Gary} {et~al.}(1975{\natexlab{b}}){Gary}, {Feldman}, {Forslund}, \&
  {Montgomery}}]{gary75b}
---. 1975{\natexlab{b}}, \jgr, 80, 4197

\bibitem[{{Gary} \& {Li}(2000)}]{gary00}
{Gary}, S.~P., \& {Li}, H. 2000, \apj, 529, 1131

\bibitem[{{Gary} \& {Saito}(2007)}]{gary07}
{Gary}, S.~P., \& {Saito}, S. 2007, \grl, 34, 14111

\bibitem[{{Gary} {et~al.}(1994){Gary}, {Scime}, {Phillips}, \&
  {Feldman}}]{gary94}
{Gary}, S.~P., {Scime}, E.~E., {Phillips}, J.~L., \& {Feldman}, W.~C. 1994,
  \jgr, 99, 23391

\bibitem[{{Gary} {et~al.}(1999){Gary}, {Skoug}, \& {Daughton}}]{gary99}
{Gary}, S.~P., {Skoug}, R.~M., \& {Daughton}, W. 1999, Phys.~Plasmas, 6, 2607

\bibitem[{{Gosling} {et~al.}(2001){Gosling}, {Skoug}, \& {Feldman}}]{gosling01}
{Gosling}, J.~T., {Skoug}, R.~M., \& {Feldman}, W.~C. 2001, \grl, 28, 4155

\bibitem[{{Graham} {et~al.}(2017){Graham}, {Rae}, {Owen}, {Walsh}, {Arridge},
  {Gilbert}, {Lewis}, {Jones}, {Forsyth}, {Coates}, \& {Waite}}]{graham17}
{Graham}, G.~A., {Rae}, I.~J., {Owen}, C.~J., {et~al.} 2017, \jgr, 122, 3858

\bibitem[{Gurgiolo \& Goldstein(2017)}]{gurgiolo17}
Gurgiolo, C., \& Goldstein, M.~L. 2017, Annales Geophysicae, 35, 71

\bibitem[{{Hammond} {et~al.}(1996){Hammond}, {Feldman}, {McComas}, {Phillips},
  \& {Forsyth}}]{hammond96}
{Hammond}, C.~M., {Feldman}, W.~C., {McComas}, D.~J., {Phillips}, J.~L., \&
  {Forsyth}, R.~J. 1996, \aap, 316, 350

\bibitem[{{He} {et~al.}(2012){He}, {Tu}, {Marsch}, \& {Yao}}]{he12}
{He}, J., {Tu}, C., {Marsch}, E., \& {Yao}, S. 2012, \apjl, 745, L8

\bibitem[{{Hollweg}(1974)}]{hollweg74}
{Hollweg}, J.~V. 1974, \jgr, 79, 3845

\bibitem[{{Horaites} {et~al.}(2018){Horaites}, {Astfalk}, {Boldyrev}, \&
  {Jenko}}]{horaites18}
{Horaites}, K., {Astfalk}, P., {Boldyrev}, S., \& {Jenko}, F. 2018, \mnras,
  480, 1499

\bibitem[{{Kennel} \& {Wong}(1967)}]{kennel67}
{Kennel}, C.~F., \& {Wong}, H.~V. 1967, J.~Plasma Phys., 1, 75

\bibitem[{{Krafft} \& {Volokitin}(2003)}]{krafft03}
{Krafft}, C., \& {Volokitin}, A. 2003, Ann.~Geophys., 21, 1393

\bibitem[{{Krafft} \& {Volokitin}(2006)}]{krafft06}
---. 2006, Phys.~Plasmas, 13, 122301

\bibitem[{{Krafft} {et~al.}(2005){Krafft}, {Volokitin}, \&
  {Zaslavsky}}]{krafft05}
{Krafft}, C., {Volokitin}, A., \& {Zaslavsky}, A. 2005, Phys.~Plasmas, 12,
  112309

\bibitem[{{Lakhina}(1977)}]{lakhina77}
{Lakhina}, G.~S. 1977, \solphys, 52, 153

\bibitem[{{Lakhina}(1979)}]{lakhina79}
---. 1979, \apss, 63, 511

\bibitem[{{Landi} \& {Pantellini}(2003)}]{landi03}
{Landi}, S., \& {Pantellini}, F. 2003, \aap, 400, 769

\bibitem[{{Lazar} {et~al.}(2013){Lazar}, {Poedts}, \& {Michno}}]{lazar13}
{Lazar}, M., {Poedts}, S., \& {Michno}, M.~J. 2013, \aap, 554, A64

\bibitem[{{Lazar} {et~al.}(2011){Lazar}, {Poedts}, \& {Schlickeiser}}]{lazar11}
{Lazar}, M., {Poedts}, S., \& {Schlickeiser}, R. 2011, Monthly Not.~Royal
  Astron.~Soc., 410, 663

\bibitem[{{Lie-Svendsen} {et~al.}(1997){Lie-Svendsen}, {Hansteen}, \&
  {Leer}}]{lie97}
{Lie-Svendsen}, {\O}., {Hansteen}, V.~H., \& {Leer}, E. 1997, \jgr, 102, 4701

\bibitem[{{Lin}(1998)}]{lin98}
{Lin}, R.~P. 1998, \ssr, 86, 61

\bibitem[{{Lin} {et~al.}(1995){Lin}, {Anderson}, {Ashford}, {Carlson},
  {Curtis}, {Ergun}, {Larson}, {McFadden}, {McCarthy}, {Parks}, {R{\`e}me},
  {Bosqued}, {Coutelier}, {Cotin}, {D'Uston}, {Wenzel}, {Sanderson}, {Henrion},
  {Ronnet}, \& {Paschmann}}]{lin95}
{Lin}, R.~P., {Anderson}, K.~A., {Ashford}, S., {et~al.} 1995, \ssr, 71, 125

\bibitem[{{Maksimovic} {et~al.}(2000){Maksimovic}, {Gary}, \&
  {Skoug}}]{maksimovic00}
{Maksimovic}, M., {Gary}, S.~P., \& {Skoug}, R.~M. 2000, \jgr, 105, 18337

\bibitem[{{Maksimovic} {et~al.}(1997){Maksimovic}, {Pierrard}, \&
  {Riley}}]{maksimovic97}
{Maksimovic}, M., {Pierrard}, V., \& {Riley}, P. 1997, \grl, 24, 1151

\bibitem[{{Maksimovic} {et~al.}(2005){Maksimovic}, {Zouganelis}, {Chaufray},
  {Issautier}, {Scime}, {Littleton}, {Marsch}, {McComas}, {Salem}, {Lin}, \&
  {Elliott}}]{maksimovic05}
{Maksimovic}, M., {Zouganelis}, I., {Chaufray}, J.-Y., {et~al.} 2005, \jgr,
  110, 9104

\bibitem[{{Marsch}(2006)}]{marsch06}
{Marsch}, E. 2006, Living Rev.~Solar Phys., 3, 1

\bibitem[{{Omelchenko} {et~al.}(1994){Omelchenko}, {Shapiro}, {Shevchenko},
  {Ashour-Abdalla}, \& {Schriver}}]{omelchenko94}
{Omelchenko}, Y.~A., {Shapiro}, V.~D., {Shevchenko}, V.~I., {Ashour-Abdalla},
  M., \& {Schriver}, D. 1994, \jgr, 99, 5965

\bibitem[{{Owens} {et~al.}(2008){Owens}, {Crooker}, \& {Schwadron}}]{owens08}
{Owens}, M.~J., {Crooker}, N.~U., \& {Schwadron}, N.~A. 2008, \jgr, 113, A11104

\bibitem[{{Pagel} {et~al.}(2005){Pagel}, {Crooker}, {Larson}, {Kahler}, \&
  {Owens}}]{pagel05}
{Pagel}, C., {Crooker}, N.~U., {Larson}, D.~E., {Kahler}, S.~W., \& {Owens},
  M.~J. 2005, \jgr, 110, 1103

\bibitem[{{Pavan} {et~al.}(2013){Pavan}, {Vi{\~n}as}, {Yoon}, {Ziebell}, \&
  {Gaelzer}}]{pavan13}
{Pavan}, J., {Vi{\~n}as}, A.~F., {Yoon}, P.~H., {Ziebell}, L.~F., \& {Gaelzer},
  R. 2013, \apjl, 769, L30

\bibitem[{{Phillips} \& {Gosling}(1990)}]{phillips90}
{Phillips}, J.~L., \& {Gosling}, J.~T. 1990, \jgr, 95, 4217

\bibitem[{{Pilipp} {et~al.}(1987{\natexlab{a}}){Pilipp}, {Muehlhaeuser},
  {Miggenrieder}, {Montgomery}, \& {Rosenbauer}}]{pilipp87}
{Pilipp}, W.~G., {Muehlhaeuser}, K.-H., {Miggenrieder}, H., {Montgomery},
  M.~D., \& {Rosenbauer}, H. 1987{\natexlab{a}}, \jgr, 92, 1075

\bibitem[{{Pilipp} {et~al.}(1987{\natexlab{b}}){Pilipp}, {Muehlhaeuser},
  {Miggenrieder}, {Rosenbauer}, \& {Schwenn}}]{pilipp87b}
{Pilipp}, W.~G., {Muehlhaeuser}, K.-H., {Miggenrieder}, H., {Rosenbauer}, H.,
  \& {Schwenn}, R. 1987{\natexlab{b}}, \jgr, 92, 1103

\bibitem[{{Pongkitiwanichakul} \& {Chandran}(2014)}]{pongkitiwanichakul14}
{Pongkitiwanichakul}, P., \& {Chandran}, B.~D.~G. 2014, \apj, 796, 45

\bibitem[{{Pulupa} {et~al.}(2014){Pulupa}, {Bale}, {Salem}, \&
  {Horaites}}]{pulupa14}
{Pulupa}, M.~P., {Bale}, S.~D., {Salem}, C., \& {Horaites}, K. 2014, \jgr, 119,
  647

\bibitem[{{Ramani} \& {Laval}(1978)}]{ramani78}
{Ramani}, A., \& {Laval}, G. 1978, Phys.~Fluids, 21, 980

\bibitem[{{Roberg-Clark} {et~al.}(2019){Roberg-Clark}, {Agapitov}, {Drake}, \&
  {Swisdak}}]{roberg19}
{Roberg-Clark}, G.~T., {Agapitov}, O.~V., {Drake}, J.~F., \& {Swisdak}, M.~M.
  2019, arXiv e-prints, arXiv:1908.06481

\bibitem[{{Roberg-Clark} {et~al.}(2016){Roberg-Clark}, {Drake}, {Reynolds}, \&
  {Swisdak}}]{roberg16}
{Roberg-Clark}, G.~T., {Drake}, J.~F., {Reynolds}, C.~S., \& {Swisdak}, M.
  2016, \apjl, 830, L9

\bibitem[{{Roberg-Clark} {et~al.}(2018){Roberg-Clark}, {Drake}, {Swisdak}, \&
  {Reynolds}}]{roberg18}
{Roberg-Clark}, G.~T., {Drake}, J.~F., {Swisdak}, M., \& {Reynolds}, C.~S.
  2018, \apj, 867, 154

\bibitem[{{Rosenbauer} {et~al.}(1977){Rosenbauer}, {Schwenn}, {Marsch},
  {Meyer}, {Miggenrieder}, {Montgomery}, {Muehlhaeuser}, {Pilipp}, {Voges}, \&
  {Zink}}]{rosenbauer77}
{Rosenbauer}, H., {Schwenn}, R., {Marsch}, E., {et~al.} 1977, J.~ Geophysics,
  42, 561

\bibitem[{{Sahraoui} {et~al.}(2012){Sahraoui}, {Belmont}, \&
  {Goldstein}}]{sahraoui12}
{Sahraoui}, F., {Belmont}, G., \& {Goldstein}, M.~L. 2012, \apj, 748, 100

\bibitem[{{Sahraoui} {et~al.}(2010){Sahraoui}, {Goldstein}, {Belmont}, {Canu},
  \& {Rezeau}}]{sahraoui10}
{Sahraoui}, F., {Goldstein}, M.~L., {Belmont}, G., {Canu}, P., \& {Rezeau}, L.
  2010, \prl, 105, 131101

\bibitem[{{Salem} {et~al.}(2003{\natexlab{a}}){Salem}, {Hoang}, {Issautier},
  {Maksimovic}, \& {Perche}}]{salem03b}
{Salem}, C., {Hoang}, S., {Issautier}, K., {Maksimovic}, M., \& {Perche}, C.
  2003{\natexlab{a}}, Adv.~Space Res., 32, 491

\bibitem[{{Salem} {et~al.}(2003{\natexlab{b}}){Salem}, {Hubert}, {Lacombe},
  {Bale}, {Mangeney}, {Larson}, \& {Lin}}]{salem03}
{Salem}, C., {Hubert}, D., {Lacombe}, C., {et~al.} 2003{\natexlab{b}}, \apj,
  585, 1147

\bibitem[{{Salem} {et~al.}(2012){Salem}, {Howes}, {Sundkvist}, {Bale},
  {Chaston}, {Chen}, \& {Mozer}}]{salem12}
{Salem}, C.~S., {Howes}, G.~G., {Sundkvist}, D., {et~al.} 2012, \apjl, 745, L9

\bibitem[{{Scime} {et~al.}(1999){Scime}, {Badeau}, \& {Littleton}}]{scime99}
{Scime}, E.~E., {Badeau}, Jr., A.~E., \& {Littleton}, J.~E. 1999, \grl, 26,
  2129

\bibitem[{{Scudder} \& {Olbert}(1979{\natexlab{a}})}]{scudder79}
{Scudder}, J.~D., \& {Olbert}, S. 1979{\natexlab{a}}, \jgr, 84, 2755

\bibitem[{{Scudder} \& {Olbert}(1979{\natexlab{b}})}]{scudder79b}
---. 1979{\natexlab{b}}, \jgr, 84, 6603

\bibitem[{{Shaaban} {et~al.}(2018{\natexlab{a}}){Shaaban}, {Lazar}, \&
  {Poedts}}]{shaaban18}
{Shaaban}, S.~M., {Lazar}, M., \& {Poedts}, S. 2018{\natexlab{a}}, \mnras, 480,
  310

\bibitem[{{Shaaban} {et~al.}(2018{\natexlab{b}}){Shaaban}, {Lazar}, {Yoon}, \&
  {Poedts}}]{shaaban18b}
{Shaaban}, S.~M., {Lazar}, M., {Yoon}, P.~H., \& {Poedts}, S.
  2018{\natexlab{b}}, Phys.~Plasmas, 25, 082105

\bibitem[{{Shaaban} {et~al.}(2019){Shaaban}, {Lazar}, {Yoon}, {Poedts}, \&
  {L{\'o}pez}}]{shaaban19}
{Shaaban}, S.~M., {Lazar}, M., {Yoon}, P.~H., {Poedts}, S., \& {L{\'o}pez},
  R.~A. 2019, \mnras, 486, 4498

\bibitem[{{Shevchenko} \& {Galinsky}(2010)}]{shevchenko10}
{Shevchenko}, V.~I., \& {Galinsky}, V.~L. 2010, Nonlinear Processes in
  Geophysics, 17, 593

\bibitem[{{Stix}(1992)}]{stix92}
{Stix}, T.~H. 1992, {Waves in plasmas} ({New York, USA}: {American Institute of
  Physics})

\bibitem[{{Tong} {et~al.}(2019){Tong}, {Vasko}, {Pulupa}, {Mozer}, {Bale},
  {Artemyev}, \& {Krasnoselskikh}}]{tong19}
{Tong}, Y., {Vasko}, I.~Y., {Pulupa}, M., {et~al.} 2019, \apj, 870, L6

\bibitem[{{{\v S}tver{\'a}k} {et~al.}(2009){{\v S}tver{\'a}k}, {Maksimovic},
  {Tr{\'a}vn{\'{\i}}{\v c}ek}, {Marsch}, {Fazakerley}, \& {Scime}}]{stverak09}
{{\v S}tver{\'a}k}, {\v S}., {Maksimovic}, M., {Tr{\'a}vn{\'{\i}}{\v c}ek},
  P.~M., {et~al.} 2009, \jgr, 114, 5104

\bibitem[{{{\v S}tver{\'a}k} {et~al.}(2008){{\v S}tver{\'a}k},
  {Tr{\'a}vn{\'{\i}}{\v c}ek}, {Maksimovic}, {Marsch}, {Fazakerley}, \&
  {Scime}}]{stverak08}
{{\v S}tver{\'a}k}, {\v S}., {Tr{\'a}vn{\'{\i}}{\v c}ek}, P., {Maksimovic}, M.,
  {et~al.} 2008, Journal of Geophysical Research (Space Physics), 113, 3103

\bibitem[{{Vasko} {et~al.}(2019){Vasko}, {Krasnoselskikh}, {Tong}, {Bale},
  {Bonnell}, \& {Mozer}}]{vasko19}
{Vasko}, I.~Y., {Krasnoselskikh}, V., {Tong}, Y., {et~al.} 2019, \apj, 871, L29

\bibitem[{{Verdon} {et~al.}(2009){Verdon}, {Cairns}, {Melrose}, \&
  {Robinson}}]{verdon09}
{Verdon}, A.~L., {Cairns}, I.~H., {Melrose}, D.~B., \& {Robinson}, P.~A. 2009,
  in IAU Symposium, Vol. 257, IAU Symposium, ed. N.~{Gopalswamy} \& D.~F.
  {Webb}, 569--573

\bibitem[{{Verscharen} {et~al.}(2013{\natexlab{a}}){Verscharen}, {Bourouaine},
  \& {Chandran}}]{verscharen13b}
{Verscharen}, D., {Bourouaine}, S., \& {Chandran}, B.~D.~G. 2013{\natexlab{a}},
  \apj, 773, 163

\bibitem[{{Verscharen} {et~al.}(2013{\natexlab{b}}){Verscharen}, {Bourouaine},
  {Chandran}, \& {Maruca}}]{verscharen13a}
{Verscharen}, D., {Bourouaine}, S., {Chandran}, B.~D.~G., \& {Maruca}, B.~A.
  2013{\natexlab{b}}, \apj, 773, 8

\bibitem[{{Verscharen} \& {Chandran}(2013)}]{verscharen13}
{Verscharen}, D., \& {Chandran}, B.~D.~G. 2013, \apj, 764, 88

\bibitem[{{Verscharen} \& {Chandran}(2018)}]{verscharen18b}
---. 2018, Res.~Notes AAS, 2, 13

\bibitem[{{Verscharen} {et~al.}(2018){Verscharen}, {Klein}, {Chandran},
  {Stevens}, {Salem}, \& {Bale}}]{verscharen18}
{Verscharen}, D., {Klein}, K.~G., {Chandran}, B.~D.~G., {et~al.} 2018,
  J.~Plasma Phys., 84, 905840403

\bibitem[{{Verscharen} {et~al.}(2019){Verscharen}, {Klein}, \&
  {Maruca}}]{verscharen19}
{Verscharen}, D., {Klein}, K.~G., \& {Maruca}, B.~A. 2019, Living Rev.~Solar
  Phys., accepted

\bibitem[{{Vocks} \& {Mann}(2003)}]{vocks03}
{Vocks}, C., \& {Mann}, G. 2003, \apj, 593, 1134

\bibitem[{{Watson}(1922)}]{watson22}
{Watson}, G.~N. 1922, {A Treatise on the Theory of Bessel Functions} ({New
  York, USA}: {Cambridge University Press})

\bibitem[{{Wilson} {et~al.}(2009){Wilson}, {Cattell}, {Kellogg}, {Goetz},
  {Kersten}, {Kasper}, {Szabo}, \& {Meziane}}]{wilson09}
{Wilson}, L.~B., I., {Cattell}, C.~A., {Kellogg}, P.~J., {et~al.} 2009, \jgr,
  114, A10106

\bibitem[{{Wilson} {et~al.}(2018){Wilson}, {Stevens}, {Kasper}, {Klein},
  {Maruca}, {Bale}, {Bowen}, {Pulupa}, \& {Salem}}]{wilson18}
{Wilson}, Lynn~B., I., {Stevens}, M.~L., {Kasper}, J.~C., {et~al.} 2018, \apjs,
  236, 41

\bibitem[{{Wilson} {et~al.}(2013){Wilson}, {Koval}, {Szabo}, {Breneman},
  {Cattell}, {Goetz}, {Kellogg}, {Kersten}, {Kasper}, {Maruca}, \&
  {Pulupa}}]{wilson13}
{Wilson}, L.~B., {Koval}, A., {Szabo}, A., {et~al.} 2013, \jgr, 118, 5

\end{thebibliography}

\end{document}